\documentclass[sn-apa]{sn-jnl}% APA Reference Style 
\usepackage{graphicx}%
\usepackage{multirow}%
\usepackage{amsmath,amssymb,amsfonts}%
\usepackage{amsthm}%
\usepackage{mathrsfs}%
\usepackage[title]{appendix}%
\usepackage{xcolor}%
\usepackage{textcomp}%
\usepackage{manyfoot}%
\usepackage{booktabs}%
\usepackage{algorithm}%
\usepackage{algorithmicx}%
\usepackage{algpseudocode}%
\usepackage{listings}%

\usepackage{lineno}

\begin{document}

%\linenumbers

\title[\textit{flimo}]{The Fixed Landscape Inference MethOd (\textit{flimo}): a versatile alternative to Approximate Bayesian Computation, faster by several orders of magnitude}


%\author[1,$\ast$]{Sylvain Moinard\ORCID{0000-0003-0201-0547}}
%\author[2]{Edouard Oudet\ORCID{0000-0001-5690-0129}}
%\author[3]{Didier Piau}
%\author[1, $\dagger$]{Eric Coissac\ORCID{0000-0001-7507-6729}}
%\author[1, $\dagger$]{Christelle Gonindard-Melodelima\ORCID{0000-0001-7945-108X}}

\author*[1]{\fnm{Sylvain} \sur{Moinard}}\email{sylvain.moinard@univ-grenoble-alpes.fr}

\author[2]{\fnm{Edouard} \sur{Oudet}}\email{edouard.oudet@univ-grenoble-alpes.fr}

\author[3]{\fnm{Didier} \sur{Piau}}\email{didier.piau@univ-grenoble-alpes.fr}

\author[1]{\fnm{Eric} \sur{Coissac}}\email{eric.coissac@metabarcoding.org}
\equalcont{These authors contributed equally to this work.}

\author[1]{\fnm{Christelle} \sur{Gonindard-Melodelima}}\email{christelle.gonindard@univ-grenoble-alpes.fr}
\equalcont{These authors contributed equally to this work.}

%\authormark{Moinard et al.}

\affil*[1]{\orgname{Univ. Grenoble Alpes, Univ. Savoie Mont Blanc, CNRS}, \orgdiv{LECA}, \postcode{FR-38000}, \orgaddress{\country{Grenoble, France}}}
\affil[2]{\orgname{Univ. Grenoble-Alpes, CNRS}, \orgdiv{Laboratoire Jean Kuntzmann (LJK)}, \postcode{FR-38000}, \orgaddress{\country{Grenoble, France}}}
\affil[3]{\orgname{Univ. Grenoble-Alpes, CNRS}, \orgdiv{Institut Fourier}, \postcode{FR-38000}, \orgaddress{\country{Grenoble, France}}}

%\corresp[$\ast$]{Corresponding author. \href{sylvain.moinard@univ-grenoble-alpes.fr}{sylvain.moinard@univ-grenoble-alpes.fr}}

%\corresp[$\dagger$]{ These authors contributed equally to this work.}


\abstract{Modelling in biology must adapt to increasingly complex and massive data. The efficiency of the inference algorithms used to estimate model parameters is therefore questioned. Many of these are based on stochastic optimization processes that require significant computing time. We introduce the Fixed Landscape Inference MethOd (\textit{flimo}), a new likelihood-free inference method for continuous state-space stochastic models. It applies deterministic gradient-based optimization algorithms to obtain a point estimate of the parameters, minimizing the difference between the data and some simulations according to some prescribed summary statistics. In this sense, it is analogous to Approximate Bayesian Computation (ABC). Like ABC, it can also provide an approximation of the distribution of the parameters. Three applications are proposed: a usual theoretical example, namely the inference of the parameters of g-and-k distributions; a population genetics problem, not so simple as it seems, namely the inference of a selective value from time series in a Wright-Fisher model; and simulations from a Ricker model, representing chaotic population dynamics. In the two first applications, the results show a drastic reduction of the computational time needed for the inference phase compared to the other methods, despite an equivalent accuracy. Even when likelihood-based methods are applicable, the simplicity and efficiency of \textit{flimo} make it a compelling alternative.  Implementations in Julia and in R are available on \url{https://metabarcoding.org/flimo}. To run \textit{flimo}, the user must simply be able to simulate data according to the chosen model.}

%The \textit{flimo} inference method is suitable to many stochastic models involving large data sets. 

%%================================%%
%% Sample for structured abstract %%
%%================================%%

% \abstract{\textbf{Purpose:} 
% 
% \textbf{Methods:} 
% 
% \textbf{Results:} 
% 
% \textbf{Conclusion:} }




\maketitle

%MSC : 62F10 : Statistics/Parametric inference/Point estimation

\section{Introduction}\label{intro}

Modelling in biology and ecology presents some important conceptual challenges, due to the increasing complexity and size of the available data. Even for the simplest models, the likelihood is often intractable, especially in population genetics problems \citep{balding_inference_2004}. Bayesian sampling methods \citep{shoemaker_bayesian_1999} are often used to study such models. In addition to the optimal solution, these methods provide the posterior distribution of the parameters. Among these, Markov chain Monte Carlo (MCMC) methods offer the advantage of yielding convergent estimators of the maximum likelihood of the data \citep{luengo_survey_2020} but at the cost of some large computational times, and, sometimes, of a complicated preliminary analysis to determine the likelihood function of the model. However, in cases where it is possible to simulate the process under study for a given set of parameters, other methods simply compare data to simulations \citep{hartig_statistical_2011, cranmer_frontier_2020}, either through summary statistics \citep{nunes_optimal_2010} or by using the full information \citep{drovandi_comparison_2022}. Approximate Bayesian Computation (ABC) methods \citep{sisson_handbook_2018} belong to this class of algorithms. The original rejection-sampling method \citep{tavare_inferring_1997, pritchard_population_1999, beaumont_approximate_2002} has been replaced by new approaches combining ABC with iterative Monte Carlo methods as in Sequential Monte Carlo ABC \citep{del_moral_adaptive_2012, dean_parameter_2014}, where the distributions are refined step by step or coupling ABC and MCMC \citep{marjoram_markov_2003, wegmann_efficient_2009}. Bayesian Synthetic Likelihood methods are another important class of simulation-based algorithms, which estimate the likelihood of summary statistics using a multivariate normal distribution \citep{wood_statistical_2010, price_bayesian_2018}. Other Bayesian methods exist, such as particle filtering \citep{fasiolo_comparison_2016}.

Some non-Bayesian methods are also used to estimate the maximum likelihood of a data set with respect to a given model. In this category, one can mention among others Expectation-Maximization algorithms \citep{dempster_maximum_1977} or forward algorithms \citep{bollback_estimation_2008} for Hidden Markov Models (HMM) . While being efficient when the model is adapted, these methods also require in general some significant computation times or a substantial preliminary theoretical analysis.
%kiefer_stochastic_1952
%and Finite Difference Stochastic Approximation methods \citep{spall_multivariate_1992}
%Stochastic Gradient Descent algorithms \citep{montavon_stochastic_2012

Like ABC methods, but out of the Bayesian framework, the Fixed Landscape Inference MethOd (\textit{flimo}), which we propose in the present paper, adjusts some summary statistics of the simulations to those of the data. But the idea behind \textit{flimo} is to replace the time-consuming rejection-sampling approach by an efficient gradient descent phase. To increase its efficiency, \textit{flimo} relies on algorithms that are usually only applicable in a deterministic framework. Many efficient local optimization algorithms for deterministic functions \citep{nocedal_numerical_2006} exist, such as quasi-Newton algorithms that require Hessian computation. However, these methods need a smooth solution landscape with a limited number of local optima since, when this condition is not met, the convergence cannot be guaranteed. This explains why these methods are not suitable for the optimization of stochastic functions. By definition, for a given set of parameters, a stochastic function may return different values. The non-constancy of the value returned during different optimization cycles induces an instability of the landscape and, thus, many spurious local optima, preventing the correct estimation of the gradient. To overcome this limitation, in \textit{flimo}, the solution landscape is stabilized by fixing the randomness of the simulation beforehand, drawing all needed random values from a unique random seed uniformly distributed on $[0, 1]$. Later on, these values will be reused for each optimization cycle, by transforming them into the appropriate distribution using quantile functions, a common approach to generate random values with a prescribed distribution. Thus, the simulations become deterministic, and the objective function to be minimized becomes stable. To our knowledge, these works are the first to use this deterministic quantile approach for random model inference problems.

To apply \textit{flimo}, all one needs to do is to simulate the process and to choose some appropriate summary statistics to compare the data and the simulations. The adaptation of existing simulators to \textit{flimo} is thus straightforward. Although \textit{flimo} was developed to provide some point estimates of the unknown parameters, it can also be used to approximate their full distributions.

To illustrate the workings of \textit{flimo}, we present three applications: a theoretical benchmark, namely the inference of the parameters of a g-and-k distribution; a population genetics problem, namely the estimation of the allelic selection parameter from some time series in a Wright-Fisher model; and the estimation of parameters of a Ricker model, that gives an example of a possible application in population dynamics. In the two first cases, we compare \textit{flimo} to other existing methods, to highlight the advantages of each approach in terms of bias and precision of estimates as well as computation time.

\section{Material and Methods}\label{sec2}

\subsection{Description of the algorithm}

The \textit{flimo} algorithm relies on the construction of a regular deterministic objective function which is built from random simulations of the process to be modelled and which is then efficiently minimized. Contrarily to classical stochastic methods, \textit{flimo} needs a single random draw to evaluate all the candidate parameters \textbf{$\theta$} and to select the best one according to the chosen summary statistics. This selection is performed using a deterministic algorithm. The steps of the \textit{flimo} algorithm are described below and summarized in Algorithm \ref{flimo}.

%From a mathematical point of view, \textit{flimo} works by fixing an outcome in the sample space $\Omega$.
%modif 28/06/23

In the studied probability space, denoted $(\Omega, \mathcal{A}, \mathbb{P})$, let $X_\textbf{$\theta$}$ be the random variable defining a drawing in the considered model, with parameters \textbf{$\theta$}. $X_\textbf{$\theta$}$ is defined as an application $X_\textbf{$\theta$} : \omega \in \Omega \mapsto X(\omega ; \textbf{$\theta$})$. From a mathematical point of view, \textit{flimo} works by fixing the outcome $\omega_0 \in \Omega$ and by considering a deterministic function called $simQ : \textbf{$\theta$} \mapsto X(\omega_0; \textbf{$\theta$})$ which gives a drawing in the model for any \textbf{$\theta$} parameters.

\begin{algorithm}
\caption{Fixed Landscape Inference MethOd}\label{flimo}
\textbf{Input:}\\
\#Data\\
\hspace*{\algorithmicindent} $y^{obs}$\\
\#Defined by model\\
\hspace*{\algorithmicindent} $n_{draw}$ \#random draws for one simulation\\
\hspace*{\algorithmicindent} $simQ$ \#adequate quantile simulator \\
\#Chosen by user\\
\hspace*{\algorithmicindent} $n_{sim}$ \#simulations to perform\\
\hspace*{\algorithmicindent} $s$ \#summary statistics\\
\hspace*{\algorithmicindent} $d$ \#distance between summary statistics \\
\textbf{Output:} $\textbf{$\widehat{\theta}$} = \mathrm{argmin} \: J$ 
\begin{algorithmic}[1]
\Statex \#Define quantiles matrix
\State $\textbf{R} \leftarrow (r_{i,j})$ with $r_{i,j} \sim \mathcal{U}([0,1])$ i.i.d.
\Statex \#Define objective function
\State $J \: : \: \textbf{$\theta$} \mapsto d(s(y^{obs}), s(simQ(\textbf{$\theta$}, \textbf{R})))$
\Statex \#Run optimization
\State $\widehat{\textbf{$\theta$}} \leftarrow \mathrm{argmin} J$
\end{algorithmic}
\end{algorithm}

\subsubsection{Preliminary drawing of the randomness}

A random simulation of a model is based on a certain number of draws of some simple random variables. For example, in the Wright-Fisher model \citep{ewens_mathematical_2004}, a binomial distribution is drawn at each generation and another one at each sampling. To run \textit{flimo}, the user needs to determine an upper bound of the number of these draws for a simulation. This value is noted $n_{draw}$. Then the user decides the number of simulations $n_{sim}$ to perform to estimate the typical summary statistics of parameters \textbf{$\theta$}. These simulations can be averaged for example. This choice is based on a trade-off between computation time and dispersion of the estimators. We suggest to test several values of $n_{sim}$, typically between 10 and 1000. Once $n_{draw}$ and $n_{sim}$ are fixed, the randomness is drawn. A matrix \textbf{$R$} of dimension $n_{sim} \times n_{draw}$ is set such that each entry $r_{i,j} \sim \mathcal{U}([0, 1])$ and the entries are independent. These values will be considered as the quantiles of each random draw involved in the $n_{sim}$ simulations and converted on purpose to a realization of the desired random distribution thanks to its quantile function parameterized using \textbf{$\theta$}.

\subsubsection{Special framework to obtain an empirical distribution}

To obtain a convenient empirical distribution, the user must set $n_{sim}$ to 1 and run several independent inferences. To improve performance, it is useful to set the initial condition of inference $n+1$ to the inferred value of inference $n$.

\subsubsection{Use of the randomness to carry out simulations}

In the chosen model with parameter \textbf{$\theta$}, the $k^{th}$ draw of the process can be written $Z_k \sim \mathcal{L}_k(\textbf{$\theta$})$ where $\mathcal{L}_k$ is a probability distribution parameterized by \textbf{$\theta$}. This draw may depend on the state of the system at step $k-1$.

The cumulative distribution function (CDF) of $\mathcal{L}_k$ is denoted by $F_k^\textbf{$\theta$}$. Recall that the quantile function $Q_k^\textbf{$\theta$}$ is defined by $Q_k^\textbf{$\theta$}(q) = \inf \{x \mid F_k^\textbf{$\theta$}(x) \geq q\}$ for every $q \in [0, 1]$. Thus, $Q^\textbf{$\theta$}_k$ acts as the inverse function of $F_k^\textbf{$\theta$}$ since $Q^\textbf{$\theta$}_k(q) \leq x$ if and only if $q \leq F^\textbf{$\theta$}_k(x)$. In the most common cases, $F^\textbf{$\theta$}_k(Q^\textbf{$\theta$}_k(q)) = q$ for every $q \in ]0, 1[$. The models considered in biology or ecology can be sophisticated but they are often based on compositions of usual distributions (normal distribution, Poisson distribution...). It is the quantile functions of these elementary blocks that must be known, which is generally verified in practice. Under \textit{flimo}, each step $Z_k \sim \mathcal{L}_k(\textbf{$\theta$})$ of simulation $i$ is replaced by equation \ref{Qflimo}.

\begin{align}
	Z_k^i = Q^\textbf{$\theta$}_k(r_{i,k}) \label{Qflimo}
\end{align}

By construction, $Z_k^i \sim \mathcal{L}_k(\textbf{$\theta$})$. Indeed, if $U \sim \mathcal{U}([0, 1])$, then $ Q^\textbf{$\theta$}_k(U)$ is a random variable with cumulative distribution function $F^\textbf{$\theta$}_k$. This method of generation of pseudo-random numbers is called inverse transform sampling. Moreover, once the matrix \textbf{$R$} is fixed, each run, and thus the whole set of simulations, becomes deterministic for any value of $\textbf{$\theta$}$. This also yields a global monotone coupling since larger position parameters of the distribution yield larger values of the drawn random variables.

Let $(\textbf{$\theta$}, \textbf{R}) \mapsto simQ(\textbf{$\theta$}, \textbf{R})$ denote the simulator using quantiles instead of random calls. It is crucial to underline that once the matrix \textbf{$R$} is fixed, $simQ(\textbf{$\theta$}, \textbf{R})$ produces exactly $n_{sim}$ independent simulations of the model with respect to $\textbf{$\theta$}$, as a classical simulator would do. We note $y^\textbf{$\theta$} = simQ(\textbf{$\theta$}, \textbf{R})$ the simulations performed for a set of parameters $\textbf{$\theta$}$.

\subsubsection{Building the objective function}

The user must then choose some summary statistics $s$, appropriate for the model studied. Many studies have been carried out on this subject \citep{nunes_optimal_2010}. As for the ABC methods, this choice plays a major role in the quality of inference. One must also choose a distance $d$ to compare the summary statistics of the data, noted $y^{obs}$, and the simulations $y^\textbf{$\theta$}$. The Euclidean norm and the Mean Absolute Deviation are two reasonable options. Once these components are chosen, the objective function $J$ is simply defined by equation \ref{objflimo}.

\begin{align}
	J(\textbf{$\theta$}) = d(s(y^{obs}), s(y^\textbf{$\theta$})) \label{objflimo}
\end{align}

The function $J$ is deterministic with the same smoothness as the implied quantile functions (again for some fixed matrix \textbf{$R$}). The usual continuous probability distributions have a smooth quantile function; on the other hand for discrete distributions the quantile function is piecewise constant.

\subsubsection{Deterministic optimization algorithm and automatic differentiation}

A deterministic local optimization algorithm is then used to estimate $\underset{\textbf{$\theta$}}{\mathrm{argmin}}\, J(\textbf{$\theta$})$. If the stochastic process involves only draws from continuous distributions (e.g normal distributions), it is possible to use a gradient-based second-order, e.g. quasi-Newton-type algorithm. In the case of discrete distributions, two routes are available. If $\textbf{$\theta$}$ is low dimensional, a gradient-free method may be suitable. Otherwise, each discrete distribution can be replaced by an adequate continuous distribution, such as a normal distribution with same mean and variance.

When the (transformed) probability distributions are continuous, the $J$ function is differentiable almost everywhere. It is then possible to accelerate the inference by using an Automatic Differentiation module \citep{bartholomew-biggs_automatic_2000, revels_forward-mode_2016}. Thus, the gradient and the Hessian of $J$ are computed automatically by \textit{chain rule} and not by finite difference which is the standard method of estimating differentials. Automatic Differentiation reduces both the risks of numerical errors and the computational times.

\subsection{Implementation}

\subsubsection{Packages overview}

Implementations of \textit{flimo} are freely available in the Julia package \texttt{Jflimo.jl} and in the R package \texttt{flimo}, both in \url{https://metabarcoding.org/flimo}. The Julia implementation takes advantage of the good numerical performances of this language \citep{bezanson_julia_2017}. This version is used for the three applications presented in the present paper. The R implementation is deposited on the CRAN. While the language R \citep{R_Core_Team_2021} is probably the data analysis language that biologists use the most, it suffers from some performance limitations when compared to Julia.

In Julia, the optimization functions come from the package \texttt{Optim.jl} \citep{k_mogensen_optim_2018}. The adequate application framework (when the objective function $J$ is differentiable) allows us to use the \textit{IPNewton} method \citep{wachter_implementation_2006}, an interior-point Newton algorithm solving constrained optimization problems. One may use this method in combination with the Automatic Differentiation module \texttt{ForwardDiff.jl} \citep{revels_forward-mode_2016}. In the case of a non-differentiable problem, the Brent method \citep{brent_algorithms_2002} is adequate in the one-dimensional case and an implementation with the Nelder-Mead optimization method \citep{gao_implementing_2012} is provided for multidimensional problems.
%nelder_simplex_1965

Our R package proposes two modes: \texttt{flimoR} and \texttt{flimoRJ}. In \texttt{flimoR} mode, \textit{flimo} is implemented with the \textit{optim} function of the R \texttt{stats} package with the L-BFGS-B method \citep{byrd_limited_1995}, a modification of the BFGS quasi-Newton method in the differentiable case. The \texttt{flimoRJ} mode uses the Julia functions implemented in \texttt{Jflimo.jl} thanks to the R package \texttt{JuliaConnectoR} \citep{Lenz_JuliaConnectoR_2022}.

\subsubsection{Building adequate simulator}

Any classical simulator of the studied process can be used, simply adapting it according to the following basic procedure. Each time random draws are performed with a random function, this must be replaced by the associated quantile function. 
To ensure the independence of the draws, each quantile must be used only once for each simulation.

In the R case, one replaces the random functions (\textit{rpois}, \textit{rnorm}...) by their quantile version (\textit{qpois}, \textit{qnorm}...). In the Julia case, the procedure is the same with the packages \texttt{Random.jl} and \texttt{Distributions.jl}: the \textit{rand} calls are replaced by some \textit{quantile} calls. In both cases, the number of random drawings has to be replaced by the adequate submatrix of \textbf{$R$}. A tutorial is provided on the git page of the project.

\subsection{Comparison to other inference algorithms}

We present in detail the two applications used to study \textit{flimo}. One can reproduce our results, using the scripts available on the git page of the project (\url{https://metabarcoding.org/flimo/flimo}). The main work was performed on a laptop MacBook Air (2017, 2.2 GHz Intel Core i7 Dual Core Processor). The results presented in supplementary material were parallelized on the GRICAD infrastructure servers.

\subsubsection{Estimate of the parameters of g-and-k distributions}

\paragraph{Definition} The family of g-and-k distributions can be viewed as an asymmetric generalization of the normal distributions $\mathcal{N}(\mu, \sigma^2)$, using two additional shape parameters $g$ and $k$. Estimating the parameters of a g-and-k distribution from a random sample is a classical example used to evaluate ABC methods \citep{sisson_handbook_2018}. The density of a general g-and-k distribution is not explicit, but its quantile function is: equation \ref{gk} holds for every $q \in [0,1]$. Here, $B > 0$, $k > -1/2$, and $z(.)$ is the quantile function of the standard normal distribution. Normal distributions are the $g=k=0$ case of this family.

\begin{align}
	Q(q \mid A, B, g, k) = A + &B \left( 1 + 0.8 \frac{1 - \exp(-gz(q))}{1 + \exp(-gz(q)} \right) \nonumber \\
    &(1+z(q)^2)^k z(q) \label{gk}
\end{align}

\paragraph{Compared inference methods} As is usually done, we choosed $\textbf{$\theta_\text{true}$} = (A, B, g, k)_\text{true} = (3, 1, 2, 0.5)$ \citep{drovandi_likelihood-free_2011}. We generated 100 independent data sets $y^{obs}$ of 1000 draws of the distribution with parameters $\textbf{$\theta_\text{true}$}$. One typical distribution is shown in Supplementary Figure S1. Other parameter sets have been studied under the same conditions.

The relative efficiencies of the MCMC, ABC and \textit{flimo} methods used with different objective functions and computational efforts were measured, our goal being to disentangle the effects of these aspects of the optimization process on the quality of the results (Table \ref{table:gkframework}). Parameter inferences are performed by each method for each of the 100 simulated data sets.

The MCMC method (available online: \url{https://github.com/pierrejacob/winference} \citep{bernton_approximate_2019}) is considered the gold standard. The Sequential Monte Carlo ABC algorithm implemented in the same package \texttt{winference} and called here \textit{wABC}, uses as summary statistics the Wasserstein distance of order 1 \citep{drovandi_comparison_2022}, thus comparing the complete collection of empirical quantiles of the two distributions $y^{obs}$ and $y^\textbf{$\theta$}$. The objective function $J$ is then defined as an average absolute deviation (equation \ref{Jwasserstein}).

\begin{align}
	J(\textbf{$\theta$})= \frac{1}{n} \sum_{i = 1}^n \vert Y^{obs}_{(i)} - Y^\textbf{$\theta$}_{(i)} \vert \label{Jwasserstein}
\end{align}

Here, $n$ denotes the common observation size and the $Y_{(i)}$ are the order statistics $Y_{(1)} \leq \dots \leq Y_{(n)}$ of the sample $Y = (Y_1, \dots, Y_n)$ under consideration.

%To ease the process, the quantiles are pre-sorted for each simulation so that the realizations are directly order statistics.
%The objective function based on this Wasserstein distance has several local minima. It was necessary to use the inferred parameters of iteration $i$ as the initial condition of iteration $i+1$ and to remove the first four inferences, used as a burn-in phase.
%wflimo : mediane des essais

This statistics has been implemented to build the so-called \textit{wflimo} method. We then explored two strategies to reduce the computation of the ABC and \textit{flimo} methods. The first method aims to reduce the computational effort of the optimization algorithm, allowing it less time to converge (\textit{wabc-short} method). The second aims to reduce the computational complexity of the objective function by replacing the Wasserstein distance with four summary statistics based on the empirical octiles \citep{drovandi_likelihood-free_2011}, which significantly reduces the amount of information used (\textit{oABC}, \textit{oflimo} and \textit{oflimo-short} methods). These summary statistics, called Moment Estimates and denoted by $s(y) = (S_A(y), S_B(y), S_g(y), S_k(y))$ for a set of realizations $y$, characterize the parameters of the distribution. With $E_i$ for $1 \leq i \leq 8$ denoting the empirical octiles of the samples, the summary statistics are defined in equations \ref{oct_stats}.

\begin{align}
    S_A &= E_4 \nonumber \\ 
    S_B &= E_6 - E_2 \nonumber \\
    S_g &= \frac{E_6 + E_2 - 2 E_4}{E_6 - E_2} \nonumber \\
    S_k &= \frac{E_7 - E_5 + E_3 - E_1}{E_6 - E_2} \label{oct_stats}
\end{align}

Following the recommendations of \citep{fearnhead_constructing_2012}, the simulation procedure is accelerated. Indeed, it is possible, using once again the inverse transform simulation method, to simulate uniform distribution order statistics $(U_{(i)})_{1 \leq i \leq 7}$ with the exponential spacings method \citep{ripley_stochastic_1987} (equation \ref{orderunif}) for a data set of size 1000 and to convert them to realizations of a g-and-k distribution with respect to $\textbf{$\theta$}$.

\begin{align}
	V_i \sim \Gamma(1000/8) \ \text{independent},\qquad U_{(i)} = \dfrac{\sum\limits_{j = 1}^i V_j}{\sum\limits_{k = 1}^7 V_k} \label{orderunif}
\end{align}

Then, the objective function $J$ chosen to evaluate the parameter sets $\textbf{$\theta$}$ is defined by equation \ref{Jgk}.

\begin{align}
	J(\textbf{$\theta$}) =&\left( \frac{S_A(y^{obs}) - S_A(y^\textbf{$\theta$})}{S_A(y^{obs})} \right)^2+ %\nonumber\\
	%&
	\left( \frac{S_B(y^{obs}) - S_B(y^\textbf{$\theta$})}{S_B(y^{obs})} \right)^2+ \nonumber\\
	&+\left( \frac{S_g(y^{obs}) - S_g(y^\textbf{$\theta$})}{S_g(y^{obs})} \right)^2+ %\nonumber\\
	%&
	\left( \frac{S_k(y^{obs}) - S_k(y^\textbf{$\theta$})}{S_k(y^{obs})} \right)^2 \label{Jgk}
\end{align}

\begin{table*}[!ht]
\begin{center}
\begin{minipage}{\textwidth}
\resizebox{0.99\textwidth}{!}{%
\begin{tabular}{ccccc} 
 \hline
 \multirow{2}{*}{Method} & Package: & Summary & Bounds & Computation \\ 
 & Implementation & statistics & or Prior & time control \\
 \hline
 \multirow{2}{*}{\textit{MCMC}} &\texttt{winference} : & \multirow{2}{*}{-} & Prior : $\mathcal{U}([0, 10]^4)$ &8000 iterations\\
 &\textit{metropolishastings}& &IC : inferred by \textit{wABC}&(burn-in : 2000)\\
 \hline
 \multirow{2}{*}{\textit{oABC}} & \texttt{gk} : &Moment& \multirow{6}{*}{Prior : $\mathcal{U}([0, 10]^4)$} & \multirow{2}{*}{$n_{sim} = 5 \times 10^6$}\\
 & 100 best simulations&Estimates&&\\ %\cline{1-3} \cline{5-5}
 \multirow{2}{*}{\textit{wABC}} & &&&$max_{time} = 180s$ \\
 &{\texttt{winference} :}& 1-Wasserstein &  & 1024 particles\\ %\cline{5-5}
 \multirow{2}{*}{\textit{wABC-short}} &{\textit{wsmc}} &distance&&$max_{time} = 18s$\\
 &&&&100 particles\\
 \hline
 \textit{oflimo} & \texttt{Jflimo} : & Moment& \multirow{2}{*}{Bounds : $[0, 10]^4$} & $n_{sim} = 1000$\\ %\cline{5-5}
 \textit{oflimo-short} & IPNewton with AD & Estimates & &$n_{sim} = 10$\\ \hline %\cline{2-3} \cline{5-5}
 \multirow{3}{*}{\textit{wflimo}} & \multirow{3}{*}{IPNewton without AD}  &1-Wasserstein & \multirow{3}{*}{IC : $\mathcal{U}([0, 10]^4)$} & $n_{sim} = 1$ \\ 
 &&distance& & best of \\
 &&&& 20 inferences\\
 \hline
\end{tabular}}
\caption{Application framework for 100 inferences of the g-and-k distribution. IC stands for Initial Condition. For each non-\textit{flimo} method, default setup are used (available online for \texttt{gk} \citep{prangle_gk_2017} and \texttt{winference} \url{https://github.com/pierrejacob/winference/blob/master/inst/tutorials/tutorial_gandk.pdf}). Initial Condition for \textit{MCMC} is the mean value of each parameter inferred by \textit{wABC}. For \textit{wABC} inferences, the running time exceeds the time limit set in parameter which is treated as a stop condition if it has been exceeded at the previous iteration.}
\label{table:gkframework}
\end{minipage}
\end{center}
\end{table*}

\paragraph{Estimation of parameters distribution}

The \textit{flimo} algorithm can also be used to approximate the distribution of the model parameters, as Bayesian methods provide posterior distributions. The \textit{MCMC} method started with the true parameters is again used as a reference and the obtained posterior marginal distributions are compared to those obtained by \textit{wABC} and \textit{wflimo}. To compare the posterior densities, a Kolmogorov-Smirnov test is used for each of the four marginal distributions of the parameters. This test is overpowered for these sample sizes, so we focused only on the test statistics $D$ and used it as a distance between distributions. Recall that $D$ is a distance of type $L^\infty$ between the CDFs, thus, for CDFs $F_1$ and $F_2$, $D$ is defined by equation \ref{KS}.

\begin{align}
	D = \sup_{x \in \mathbb{R}}\vert F_1(x) - F_2(x) \vert \label{KS}
\end{align}

\subsubsection{Estimate of the selection value in a Wright-Fisher model}

\paragraph{Model definition} \textit{Flimo} has then been applied to infer the strength of selection in a Wright-Fisher model \citep{ewens_mathematical_2004} from some time series of allele frequencies \citep{paris_inference_2019}. This is a typical study of population genetics problems. The distribution of two alleles $A_0$ and $A_1$ of a locus in a population of size $N_e$ is simulated over several generations, with a selective value $s$ applied on the allele $A_1$. At regular intervals, the allele frequencies are estimated by sampling a part of the population. Noting $X{(t)}$ the actual proportion of the alternative allele $A_1$ at generation $t$, the model is written with a binomial distribution (equations \ref{WF} and \ref{dff}). The function $f$ is defined on $[0, 1] $, where $s \geq -1$ is the selective value of $A_1$ and $h \in [0, 1]$ is the dominance parameter.

\begin{align}
	X{(t+1)} \mid X{(t)} &\sim \frac{1}{N_e} \mathcal{B}\left(N_e, f\left( X{(t)} \right) \right) \label{WF} \\
f(x) &= \frac{x(1+sh+s(1-h)x)}{1+2shx+s(1-2h)x^2} 
	\label{dff}
\end{align}
The available data is sampled at times $t_1 = 0 < t_2 < \dots < t_n = T$. Let $X_k = X{(t_k)}$. At each observation time, $n_k$ alleles ($n_k / 2$ individuals) are sampled. The number of $A_1$ alleles sampled is denoted by $Y_k$, hence, conditionally on $X_k$, the distribution of $Y_k$ is binomial (equation \ref{WFsample}).

\begin{align}
	Y_k \mid X_k \sim \mathcal{B}\left(n_k, X_k \right) \label{WFsample}
\end{align}

\paragraph{Approach} We compared the \textit{flimo} method with the \textit{compareHMM} method \citep{paris_inference_2019}. Thanks to the different approximations implemented in its model, this method is one of the fastest available today, while having a higher accuracy than e.g. the \textit{WFABC} method \citep{foll_wfabc_2015} used in the same context. \textit{CompareHMM} is non-Bayesian and relies on Maximum Likelihood Estimation. It follows a previous approach \citep{bollback_estimation_2008} which uses a forward algorithm on the Hidden Markov Model to compute the likelihood of the model with either the exact binomial model (\textit{compareHMM-Bin}) or an approximate model using a Beta with spikes (\textit{compareHMM-Bws}) distribution \citep{tataru_inference_2015} where transitions from one generation to the next are represented by a mixture model with a probability that the allele frequency is fixed at 0 or 1, and a Beta distribution conditional on non-fixation otherwise. \textit{CompareHMM-Bin} has an algorithmic complexity of $\mathcal{O}(N_e^3)$ so its computation time makes it prohibitive for large populations. It was used as a reference for $N_e =10^2$ and $N_e=10^3$. The results of \citep{paris_inference_2019} have been reproduced from the python code available online (\url{https://github.com/CyrielParis/compareHMM/}).

%The values of $s$ to be tested are set from $-1$ to $1$ with 300 points with the step used in \citep{paris_inference_2019}.

For \textit{flimo}, the unknown initial value of $X{(0)}$ is estimated by $\widehat{X{(0)}} = \dfrac{Y_0}{n_0}$. The unknown parameter is then just the $A_1$ selective value $\theta = s$ while the other quantities are assumed to be known. The objective to minimize is the mean absolute deviation around the median, as defined in equation \ref{JMAD}.

\begin{align}
	J(\textbf{$\theta$}) = Mean\left(\vert Y_k^\text{obs} - Median(Y_k(\textbf{$\theta$}))\vert \right)_{1\leq k \leq n} + 10^{-2} \, \vert \theta \vert \label{JMAD}
\end{align}

%We also added a safety: if one reaches a minimum larger than 0.15, another inference is launched to prevent some false convergence due to local minima.

%Hereby, every transition was simulated, contrary to the implementation of \citep{paris_inference_2019} which consists in computing by recurrence approximate moments of this distribution. 

Here, $Median(Y_k(\textbf{$\theta$}))$ is the median of the $n_{sim}$ simulations done with parameter $\textbf{$\theta$}$. The correction term is present to avoid wrong convergence results because $J$ is almost constant over a wide interval close to the lower bound. We used three different adaptations of the Wright-Fisher model. The first model is the Beta with spikes (Bws) approximated model as in \textit{compareHMM-Bws} used with a gradient-based optimization. This model does not work well with \textit{flimo} as the quantiles of the Beta distribution have no closed analytical form. One needs to inverse numerically the CDF $F(x;\alpha, \beta)$, a step which takes a substantial time. The samplings are simulated with approximated normal distributions instead of binomial distributions. The second model is the Nicholson Gaussian (NG) approximated model \citep{nicholson_assessing_2002} with the same optimization process. Classically, each binomial distribution is replaced by the normal distribution with same mean and variance, with absorbing states at allele frequencies 0 and 1. The third model is the original binomial model with a gradient-free optimization. The number of simulations $n_{sim}$ (10 and 200) was chosen for a compromise between efficiency and robustness. The exact model is more difficult to optimize due to its piecewise constant objective function. This is why it leads to worse performances and the approximation of the model is relevant.

\paragraph{Simulated data} We considered populations with three effective sizes $N_e= 10^2$, $10^3$, and $10^4$ and the initial proportion $X(0) = 0.2$. The value $N_e = 10^2$ allows to test the robustness of the inferences despite the strong stochastic variations linked to the small population size, while $N_e=10^4$ provides a very robust data set. In accordance with a range of tests performed in \citep{paris_inference_2019}, we set the dominance parameter to $h = 0.5$ and we simulated over $T = 45$ generations with a sampling every $\Delta t = 5$ generations of $n_k = 0.3 N_e$ alleles. In the two first scenarios, 100 data sets were simulated to compare the performance of the \textit{flimo} method with \cite{paris_inference_2019}. For the main scenario studied, declined in two cases, $N_e = 10^2$ and $N_e = 10^3$, the selective value is set to $s = 0.1$ (slight selective advantage for $A_1$). Data is shown in Supplementary Figure S3. Then we used a single data set with $N_e = 10^4$ and $s = 0.1$ to study the influence of the number of simulations in the dispersion of the estimate, with $n_{sim}$ chosen in 10, 20, 50, 100, 200, 500 and 1000. Two extreme scenarios were also analyzed with $N_e = 10^3$, and with $s = 0.01$ and $s = 1$.

\subsubsection{Chaotic population dynamics under Ricker model}

\begin{figure}[!ht]
 \centering
 \includegraphics[width=0.8\linewidth]{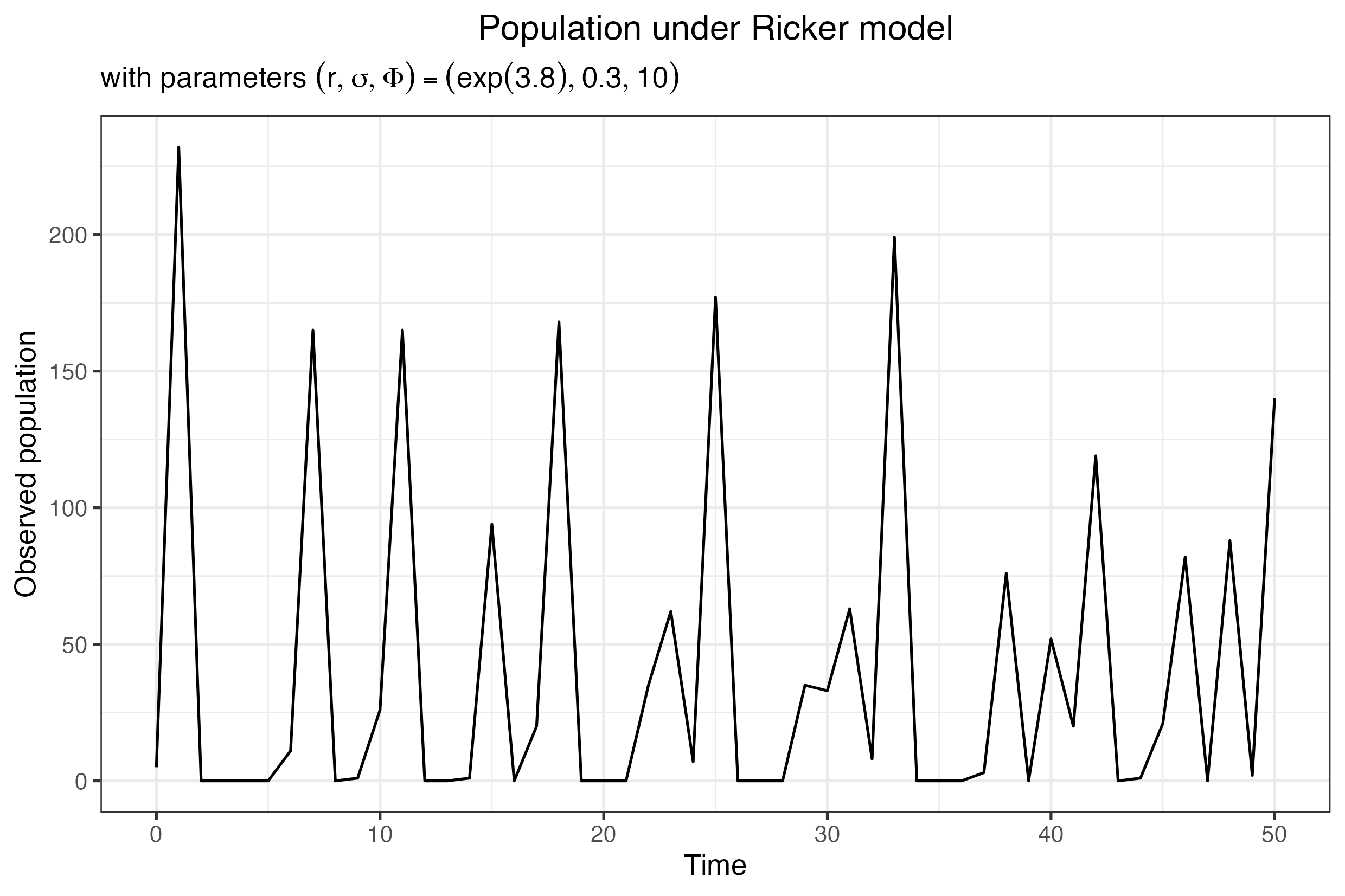}
 \caption{Typical population simulated from Ricker model with parameters\\ $(r, \sigma, \Phi) = (exp(3.8), 0.3, 10)$ and $T=50$.}
 \label{fig:ricker_sim}
\end{figure}

\textit{Flimo} is also applicable to more chaotic models. We present here an application based on the Ricker model which can describe erratic population dynamics. This model is known for the high non-linearity of its likelihood \citep{wood_statistical_2010}. This model is written according to equation \ref{ricker}. The observable data are $(Y_t)_{0\leq t \leq T}$ for a given time limit $T$. It has three parameters: $\theta = (r, \sigma, \Phi) \in \mathbb{R}_+^3$. A typical realisation of this model is shown in Figure \ref{fig:ricker_sim}.

\begin{align}
    Y_t &\sim Poisson(\Phi N_t) \nonumber \\
    N_{t+1} &= r N_t \: e^{-N_t+e_t} \label{ricker}\\
    \text{with } e_t &\sim \mathcal{N}(0, \sigma^2) \: iid \text{ and } N_0 = 1 \nonumber
\end{align}

The inference methods developed for this model fall into two main categories: an information-reduction approach (ABC or synthetik likelihood), or state space methods (particle MCMC, iterated filtering, parameter cascading) \citep{fasiolo_comparison_2016}. We used \cite{wood_statistical_2010}'s work to develop an inference protocol adapted to \textit{flimo}. The summary statistics he used were studied and six of those that vary most regularly with the parameters, in the context of random simulations carried out by fixing the randomness beforehand were kept ($s_1, \dots, s_6$). Those are the 4 parameters of the cubic regression of the ordered $Y_{t+1} - Y_t$ values, the average population and the number of observed zeros. The objective function to be minimized is given by equation \ref{J_ricker}.

\begin{align}
    J(\theta) = \sum_{k=1}^6 \left(\frac{s_k(simQ(\theta))-s_k(data)}{s_k(data)}\right)^2 \label{J_ricker}
\end{align}

 \textit{Flimo} was applied to 100 simulated data sets for seven different parameter sets, around the values $(r, \sigma, \Phi) = (exp(3.8), 0.3, 10)$ studied by \cite{wood_statistical_2010}. The optimization method chosen is the Nelder-Mead gradient-free method, with either 100 or 500 (as in \cite{wood_statistical_2010}) simulations used by \textit{flimo}.

\section{Results}\label{results}

\subsection{Parameter inference for g-and-k distributions}

\subsubsection{Point estimate of the parameters}

\begin{figure}[!ht]
 \centering
 \includegraphics[width=\linewidth]{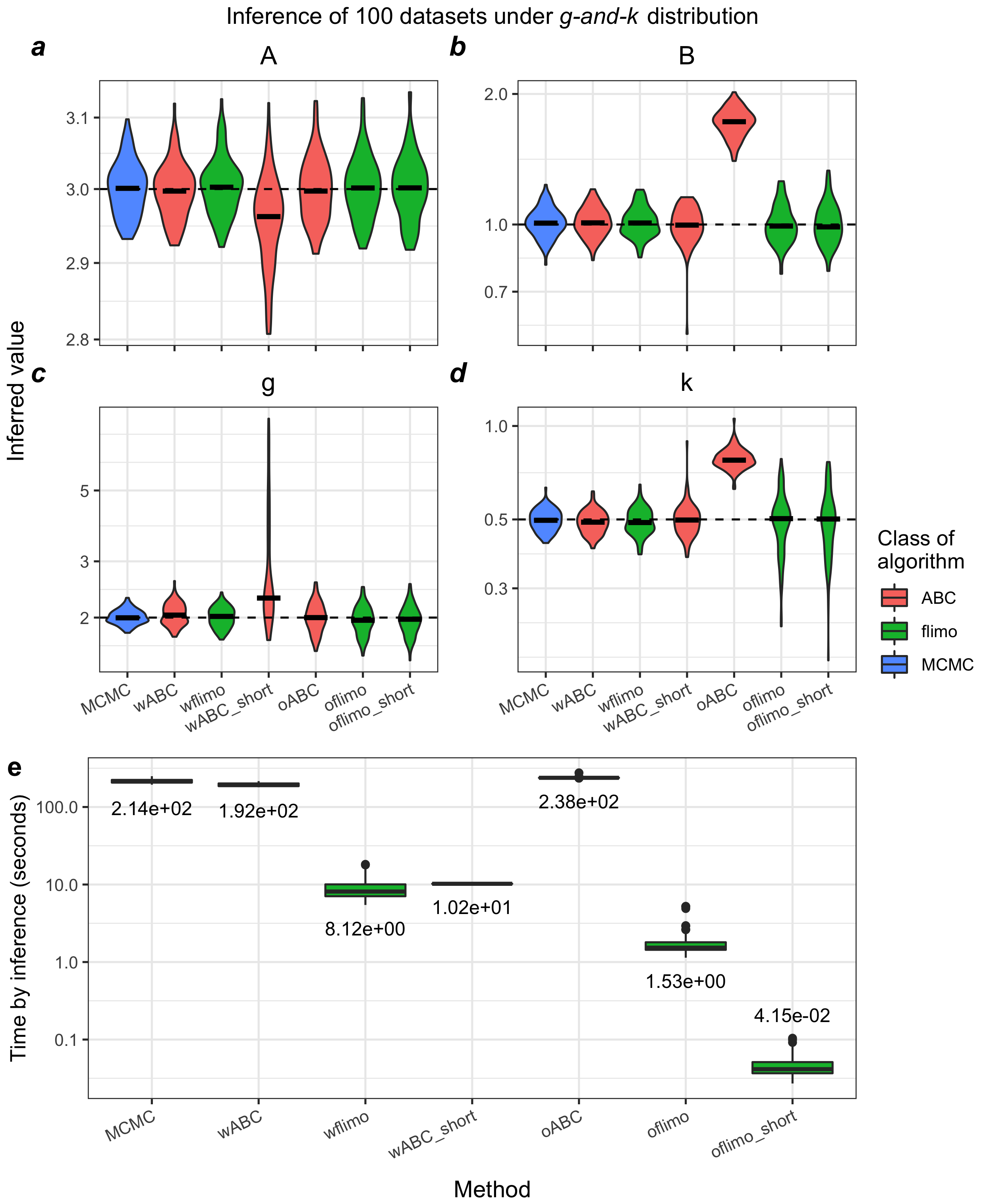}
 \caption{Inference of the parameters of g-and-k distributions for 100 independent data sets of 1000 observations and different methods (see Table \ref{table:gkframework}). Each y-scale is logarithmic. \textbf{Panels a-d}: Inference results for each parameter. Medians are plotted as bold black lines. Horizontal dashed lines correspond to the simulated values of the parameters. \textbf{Panel e}: Box plot of the 100 running times in seconds.}
 \label{fig:gkboxplots}
\end{figure}

The \textit{wABC} and \textit{wflimo} methods provide estimates of all four parameters of g-and-k distributions, which are consistent with those obtained by \textit{MCMC}. The averages of the estimates over the 100 simulated data sets yield p-values greater than 0.09 for all the different optimization methods hence, according to Wilcoxon tests with the usual significance level 0.05, they are not statistically distinguishable (Figure \ref{fig:gkboxplots}a-d).

The accuracy of the three methods is also comparable (Figure \ref{fig:gkboxplots}a-d). For $A$ and $B$, the variances of the estimates are not significantly different. When comparing \textit{wABC} to \textit{MCMC}, the variance is multiplied by a factor of 1.06 for $A$ and by a factor of 0.99 for $B$ (the p-values of Fisher tests being 0.756 and 0.984, respectively). For the comparison between \textit{wflimo} and \textit{MCMC}, the increase factors are 1.14 and 1.03, respectively (with p-values 0.522 and 0.898). The shape parameters are estimated with less precision. To wit, the estimates of $g$ are significantly more dispersed with the \textit{wABC} and \textit{wflimo} methods than with \textit{MCMC}, with a variance ratio to \textit{MCMC} of 2.38 and 1.97, respectively (p-values $<$ 0.002 for both methods). The dispersion of the estimates of $k$ is not significantly different between \textit{wABC} or \textit{wflimo}, and \textit{MCMC}, the respective variance ratio 1.10 and 1.50 has a p-value close to the 5\% threshold for the \textit{wflimo} versus \textit{MCMC} comparison (p-value 0.060). Finally, while no significant difference can be shown between the three methods in terms of bias and accuracy, \textit{wflimo} runs 26.4 times faster than \textit{MCMC} and 23.6 times faster than \textit{wABC} (Figure \ref{fig:gkboxplots}e).

When the \textit{wABC} parameters are adjusted to achieve a computation time comparable to \textit{wflimo} (\textit{wABC-short}), the method ceases to be reliable and strong biases on the estimates of the parameters $A$ and $g$ appear (Figure \ref{fig:gkboxplots}a and c), as well as an increase in the variance for each parameter compared to \textit{wABC} (with variance ratios 2.58 for $A$, 1.38 for $B$, 83.7 for $g$, and 2.39 for $k$). If the second strategy, based on a less complex objective function, is applied to the ABC method (\textit{oABC}), it allows a correct estimation of the parameters $A$ and $g$ but the estimates of $B$ and $k$ become strongly biased. It is also less efficient in terms of computation time (Figure \ref{fig:gkboxplots}e). As regards the effects of these methods of reduction of the computational times, \textit{flimo} is more robust. The use of the simplified objective function (\textit{oflimo}) reduces the computational time compared to \textit{wflimo} by a factor of 5.3 without introducing any bias on the estimates (the p-values comparing the estimates to the simulated parameters are 0.631, 0.139, 0.666, and 0.143). Only the variance of the estimates increases relative to \textit{wflimo}, by ratios of 1.14, 1.67, 1.84, 4.72 respectively (with p-values 0.5, 0.01, 0.003, $<0.001$ respectively). When the computational effort reduction is applied in conjunction with \textit{oflimo} (\textit{oflimo-short}), no significant estimation bias occurs (p-values: 0.803, 0.963, 0.758, 0. 497) and no significant increase in the variances of the estimates compared to \textit{oflimo} is observable (variance ratios: 1.10, 1.10, 1.05, 1.18; p-values: 0.632, 0.622, 0.880, 0.448). This last method, combining the two optimization procedures, runs 5157 times faster than \textit{MCMC}, without introducing any significant bias on the parameters estimation and only increasing the variance by a factor of 1.45, and 1.91 for $A$ and $B$, respectively. In contrast, for the two shape parameters, the increase in variance is much larger: 3.79 and 8.36 for $g$ and $k$, respectively.
%, see Supplementary Fig. \ref{fig:stdnsim})

The inference results for other parameter sets, with comparable conclusions, are shown in Supplementary Figure S2. The time have not been included, due to different parallelisation processes between the different methods.

The different versions of \textit{flimo} were compared for the \textit{oflimo-short} method. The \texttt{flimoR} version is about 30 times slower than \texttt{Jflimo.jl} (median time of 0.50s instead of 0.016s for one inference). The \texttt{flimoRJ} mode has a similar computation time to \texttt{Jflimo.jl} with an additional fixed cost of about 2s corresponding to the switch from R to Julia, regardless of the number of inferences.

\subsubsection{Estimation of the parameters distributions}

\begin{figure}[!ht]
 \centering
 \includegraphics[width=0.99\linewidth]{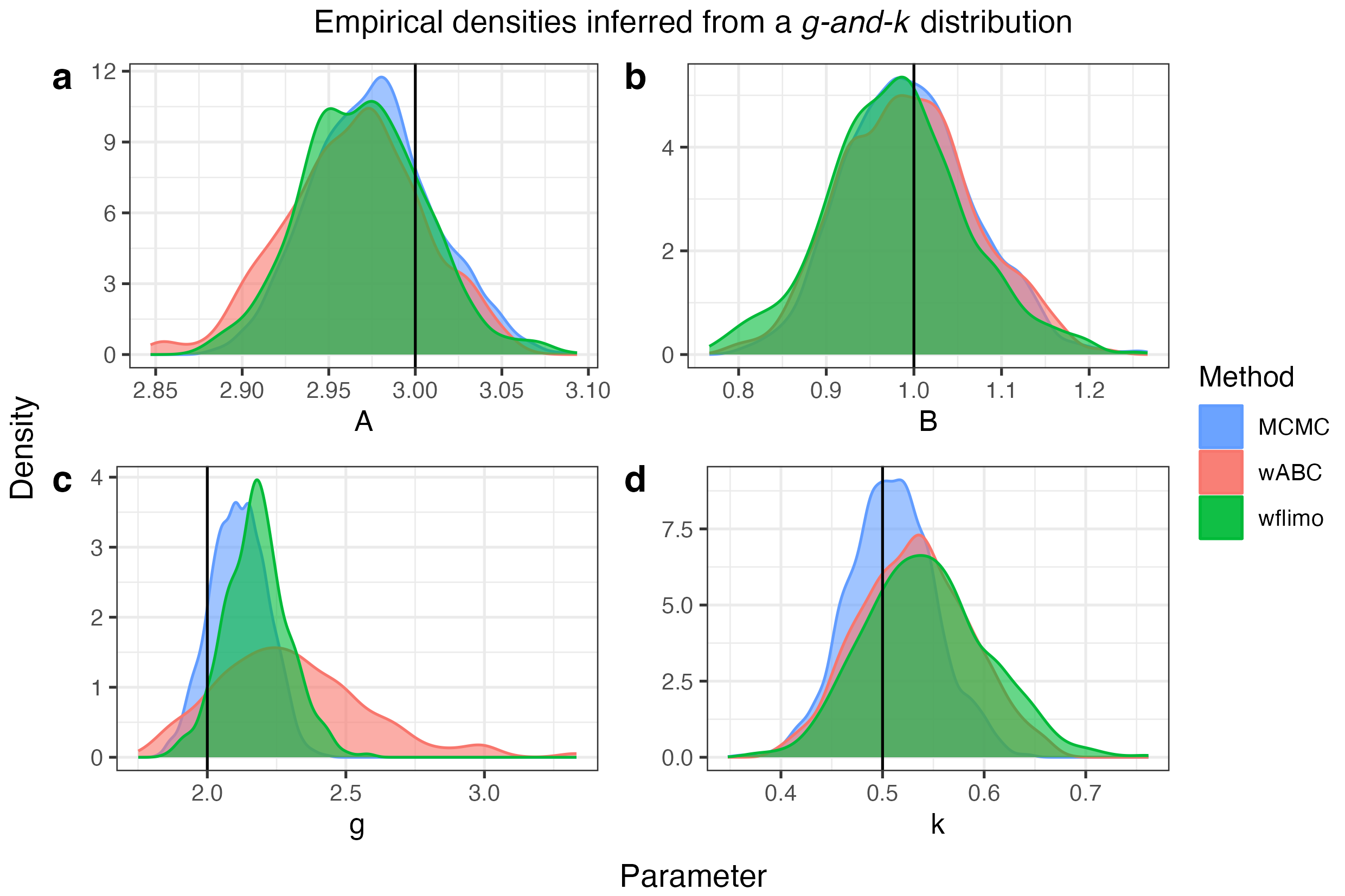}
 \caption{The empirical distributions of the four parameters of g-and-k distributions as estimated by different methods. The estimation of the distributions by the method \textit{MCMC}, used as reference (in grey), is compared to two approximations, the first one provided by the \textit{wABC} method (in blue), the second one by the \textit{wflimo} method (in red).}
 \label{fig:gkdensity}
\end{figure}

The quantile function of the density obtained by \textit{wflimo} is comparable to those obtained by the \textit{MCMC} and \textit{wABC} methods. (Figure \ref{fig:gkdensity}a-d). For the parameters $A$ and $B$, the three densities are closely related: $D$ (equation \ref{KS}) equals 0.076 versus 0.11 for $A$, 0.082 versus 0.039 for $B$, for the comparisons \textit{wflimo} versus \textit{MCMC} and \textit{wABC} versus \textit{MCMC}. For $g$, the distribution obtained by \textit{wflimo} is closer to that obtained by \textit{MCMC} than by \textit{wABC} (0.25 versus 0.45). For $k$, the densities obtained by \textit{wflimo} and \textit{wABC} are equivalent (0.28 versus 0.21). However, the current implementation of \textit{flimo} is not optimized to work in such a mode. Indeed, \textit{wflimo} takes 252s to perform 1024 inferences, while \textit{wABC} takes only 192s.

\subsection{Estimation of a Wright-Fisher selection value}

\begin{figure}[!ht]
 \centering
 \includegraphics[width=\linewidth]{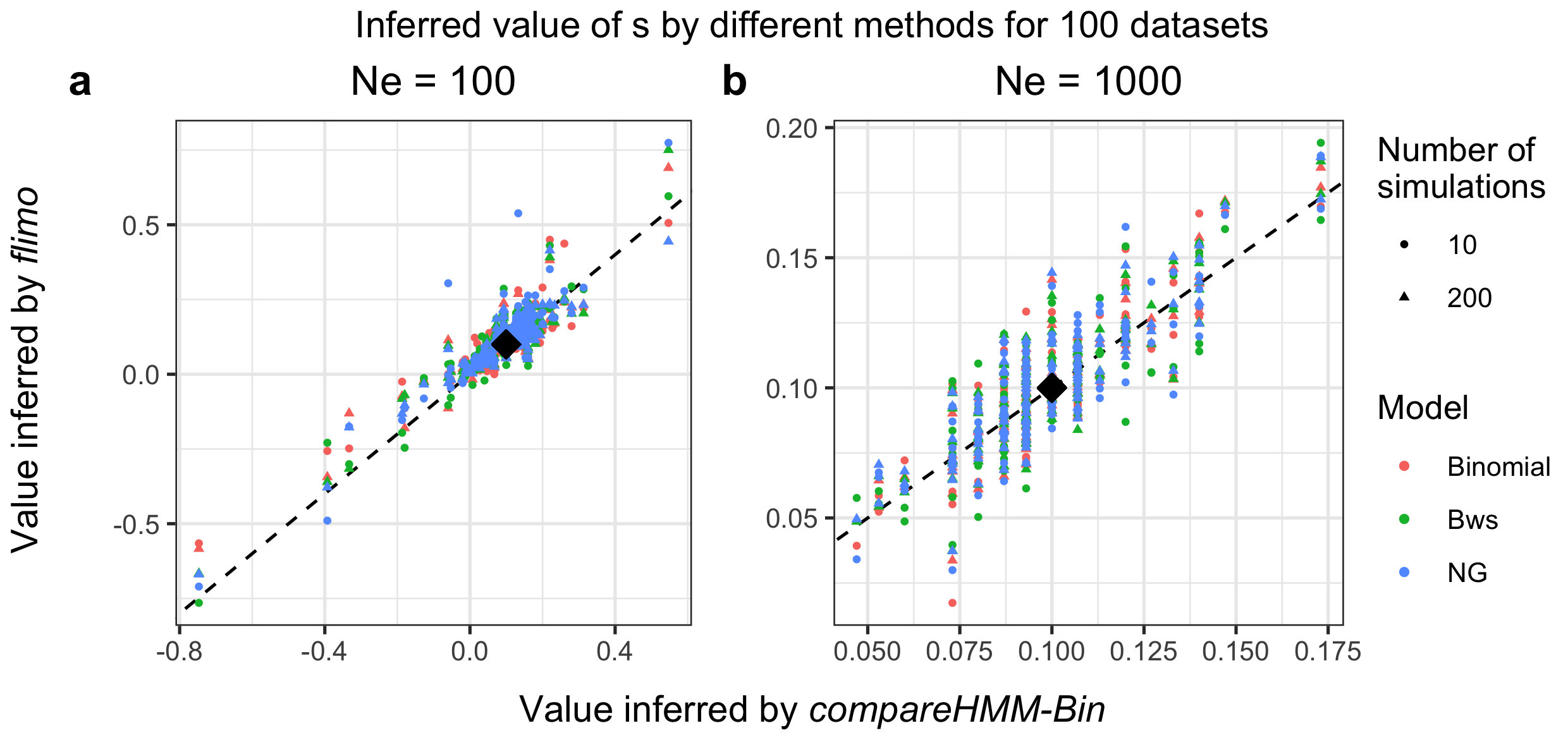}
 \includegraphics[width=\linewidth]{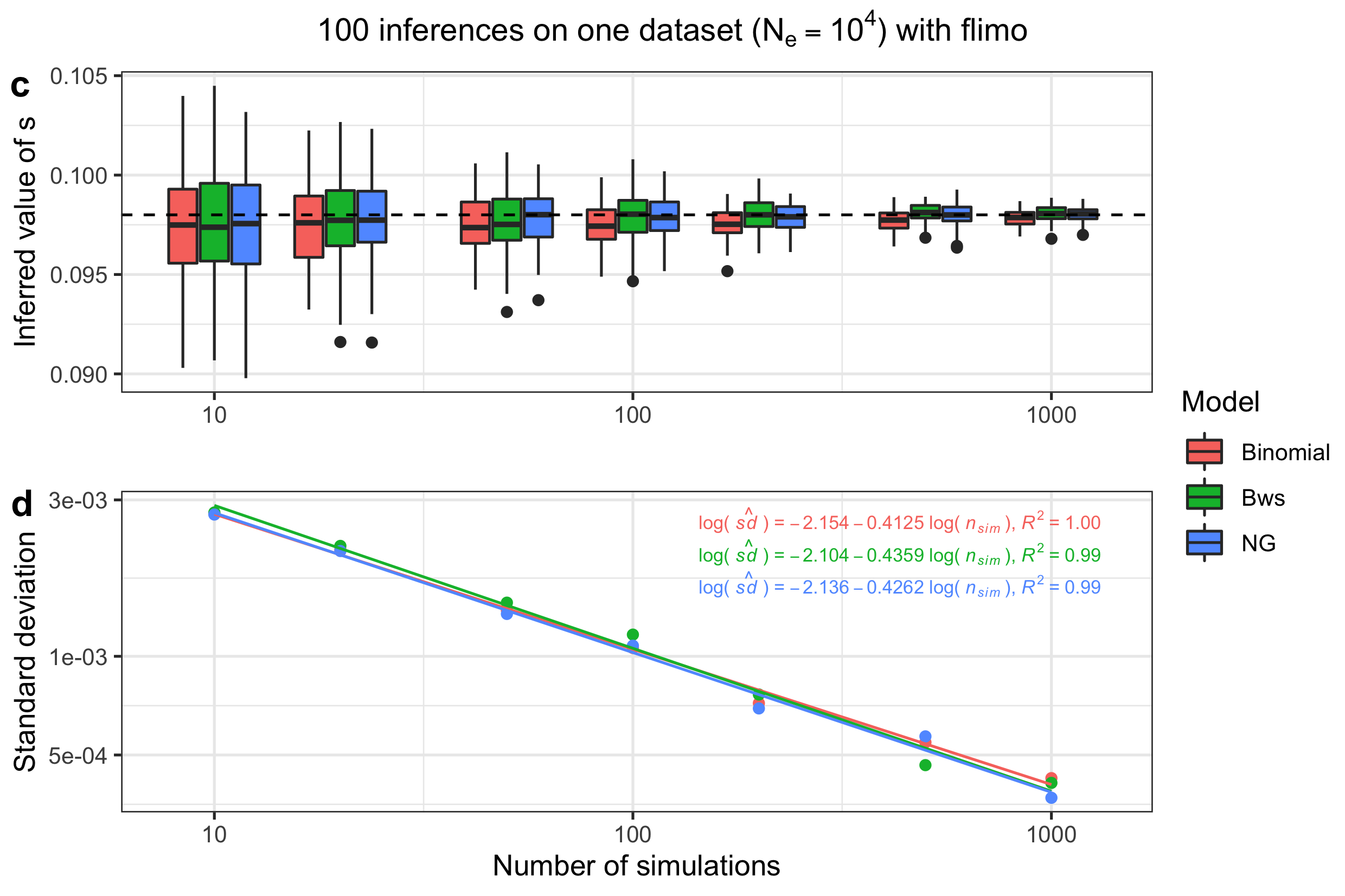}
 \caption{Comparison of inferred selection values $s$ by the \textit{compareHMM} method and different implementations of \textit{flimo}. \textbf{Panels a and b}: compared inferred values for one hundred data sets with effective population size $N_e = 10^2$ (\textbf{a}) and $N_e = 10^3$ (\textbf{b}). Dashed line corresponds to identity. Black diamond is the simulated value $s = 0.1$. \textbf{Panels c and d}: Influence of the number of simulations on \textit{flimo} inferences for a population of size $N_e = 10^4$. One hundred inferences were run for every number of simulations. \textbf{Panel c}: inferred values of $s$. The dashed line represents the value inferred by \textit{compareHMM-Bws}, $\widehat{s} = 0.098$. \textbf{Panel d}: standard deviation of the inferred values of $s$ with linear regression curve.}
 \label{fig:boxplotsWF}
\end{figure}

For $s=0.1$ and for $N_e = 10^2$ and $N_e = 10^3$, both methods exhibit highly correlated $s$ values over the 100 simulated data sets (namely, $R \in [0.88,94]$ for $N_e=10^2$ and $R \in [0.82,87]$ for $N_e=10^3$, Figure \ref{fig:boxplotsWF}a-b and Table \ref{table:wf}). Compared to \textit{compareHMM-Bin}, for $N_e=10^2$, \textit{flimo} systematically overestimated $s$ by about $10-15\%$. This systematic over-estimation almost disappears for $N_e=10^3$, with a difference of estimation between the methods ranging from $0.03\%$ to $3\%$ (Table \ref{table:wf}). For $N_e = 10^2$, the inferred mean value of $s$ is 0.883 for \textit{compareHMM-Bin} and 0.896 for \textit{compareHMM-Bws} versus $\widehat{s} \in [0.104, 0.114 ]$ for the different implementations of \textit{flimo}. For $N_e = 10^3$, we have $\widehat{s}_{compareHMM-Bin} = 0.0991$, $\widehat{s}_{compareHMM-Bws} = 0.992$ and $\widehat{s}_{flimo} \in [0.1001, 0.1013 ]$. This shows that the values inferred by \textit{flimo} are not further from the expected value $s = 0.1$ than the values inferred by \textit{compareHMM}. On average, an inference by \textit{compareHMM-Bin} lasted 1.2s for $N_e = 10^2$ and $1.0\: 10^3$s for $N_e = 10^3$.

For $N_e=10^4$, a single population was simulated. As expected, the differences between the $s$ values estimated using \textit{flimo} and \textit{compareHMM-Bws} are very small (less than $1\%$) even with $n_{sim} = 10$, a very small number of simulations. The only noticeable effect of increasing the number of simulations used by the \textit{flimo} algorithm is to reduce the standard deviation of the estimates by a factor close to $10^{-2}/\sqrt{n_{sim}}$ (Figure \ref{fig:boxplotsWF}c). For the two extreme scenarios $s= 0.01$ and $s= 1$, \textit{flimo} behaves similarly to what was observed for $s=0.1$ (Supplementary Table S1).

\begin{table*}[ht]
\begin{center}
\begin{minipage}{\textwidth}
\resizebox{\textwidth}{!}{%
\begin{tabular}{ccccccccc}
\toprule%
 Population & & \textit{compareHMM} & \multicolumn{6}{@{}c@{}}{\textit{flimo}} \\ \cmidrule{4-9}
size & Criterion & \textit{Bws} & \multicolumn{2}{@{}c@{}}{\textit{Binomial}} & \multicolumn{2}{@{}c@{}}{\textit{Bws}}& \multicolumn{2}{@{}c@{}}{\textit{NG}}\\ \cmidrule{4-5} \cmidrule{6-7} \cmidrule{8-9}
&&&10& 200 & 10& 200 & 10 & 200 \\
 \hline
 & Correlation &0.9994&0.89&0.91&0.93&0.93&0.88 & 0.94\\
\cmidrule{2-5} \cmidrule{6-7} \cmidrule{8-9}
 \multirow{5}{*}{$N_e = 10^2$} & Median difference & &\multirow{2}{*}{$0.011$}  & \multirow{2}{*}{$0.011$} & \multirow{2}{*}{$0.0096$} & \multirow{2}{*}{$0.014$} & \multirow{2}{*}{$0.016$} & \multirow{2}{*}{$0.016$} \\
 &$\widehat{s}_\text{flimo} - \widehat{s}_\text{compareHMM-Bws}$ & & & & & & & \\ 
\cmidrule{2-5} \cmidrule{6-7} \cmidrule{8-9}
 &Seconds by &\multirow{2}{*}{1.4} & \multirow{2}{*}{0.016} &\multirow{2}{*}{0.29} & \multirow{2}{*}{0.17} &\multirow{2}{*}{3.3} &\multirow{2}{*}{0.014} &\multirow{2}{*}{0.11}\\
&inference &&&&&&&\\
\hline
 &Correlation &0.9997&0.85&0.87&0.82&0.88&0.84 & 0.87\\ 
 \cmidrule{2-5} \cmidrule{6-7} \cmidrule{8-9}
 \multirow{4}{*}{$N_e = 10^3$} &Median difference & &\multirow{2}{*}{$2.5\:10^{-3}$} & \multirow{2}{*}{$4.5\:10^{-4}$} & \multirow{2}{*}{$2.2\:10^{-4}$} & \multirow{2}{*}{$1.5\:10^{-3}$} & \multirow{2}{*}{$8.8\:10^{-3}$} & \multirow{2}{*}{$1.3\:10^{-3}$} \\ 
 &$\widehat{s}_\text{flimo} - \widehat{s}_\text{compareHMM-Bws}$ & && & & & & \\ 
  \cmidrule{2-5} \cmidrule{6-7} \cmidrule{8-9}
&Seconds by &\multirow{2}{*}{1.4} & \multirow{2}{*}{0.037} &\multirow{2}{*}{0.69} & \multirow{2}{*}{0.34} & \multirow{2}{*}{7.1} &\multirow{2}{*}{0.013} &\multirow{2}{*}{0.10} \\
&inference &&&&&&&\\
 \hline
\end{tabular}}
\caption{Inference results, based on 100 simulated data sets, using \textit{compareHMM-Bin}, \textit{compareHMM-Bws} or \textit{flimo} for two different numbers of simulations and three versions of the Wright-Fisher model. Three quantities are presented: the Pearson correlation coefficient, the median of the difference between the values inferred by \textit{compareHMM-Bin} and the other methods, and the median of the computation times.}
\label{table:wf}
\end{minipage}
\end{center}
\end{table*}

%(supplementary Table \ref{table:wf1e4})

\subsection{Chaotic population dynamics under Ricker model}

The inference results are shown in Table \ref{table:ricker}. The computation time for one inference is of the order of 15-20 seconds for $n_{sim}=100$ and of 80-90 seconds for $n_{sim}=500$ (the number used in \cite{wood_statistical_2010}). The estimators are globally unbiased, with a rather large dispersion. In the case where the parameter $r$ is maximal (89), the inference fails: it might be necessary to include other summary statistics to deal with this extreme case. It is often necessary to run several inferences with different starting conditions, because of bad convergences that are detected by their high error score. Some remain in the results after 20 tries, especially in scenarios where the parameters $r$ or $\phi$ are lower. These have a large impact on the standard deviation of the estimators. The passage from $n_{sim}=100$ to $n_{sim}=500$ does not seem to improve significantly the accuracy of the inference: increasing the number of possible tries rather than the number of simulations might be relevant.

The interest of \textit{flimo} in this framework is that it is not necessary to consider the likelihood of the model, nor to evaluate the likelihood of the summary statistics as the synthetic likelihood approach do. Neither is it necessary to approximate the model to achieve a reasonable point inference of the parameters. The choice of suitable summary statistics is again crucial here: \cite{wood_statistical_2010}'s study has been very useful for our work.

\begin{table*}[ht]
\begin{center}
\begin{minipage}{\textwidth}
\resizebox{\textwidth}{!}{%
\begin{tabular}{cccccccc}
\toprule%
 \multirow{2}{*}{$r^*$} & \multirow{2}{*}{$\sigma^*$} & \multirow{2}{*}{$\Phi^*$} & \multirow{2}{*}{$n_{sim}$} & Median inferred $r$  & Median inferred $\sigma$ & Median inferred $\Phi$& Detected\\
 &&&&($1 s.d.$)&($1 s.d.$)&($1 s.d.$)&Outliers\\
  \hline
  \multirow{2}{*}{$\simeq 45$} & \multirow{2}{*}{0.3} & \multirow{2}{*}{10} & 100 & $44 \:(8.5)$ & $0.30  \:(0.13)$ & $10  \:(0.56)$ & 0\\
  &&& 500 & $45  \:(8.8)$ & $0.27  \:(0.14)$ & $10  \:(0.58)$ & 0\\
  \hline
  \multirow{2}{*}{$\simeq 22$} & \multirow{2}{*}{0.3} & \multirow{2}{*}{10} & 100 & $21  \:(7.4)$ & $0.31  \:(0.14)$ & $10  \:(23)$ & 12\\
  &&& 500 & $21  (\:7.9)$ & $0.30  \:(0.15)$ & $10  \:(25)$ & 15 \\
  \hline
  \multirow{2}{*}{$\simeq 89$} & \multirow{2}{*}{0.3} & \multirow{2}{*}{10} & 100 & $82  \:(17)$ & $0.41  \:(0.20)$ & $10  \:(0.87)$ & 0\\
  &&& 500 & $85  \:(15) $ & $0.35  \:(0.18)$ & $10  \:(0.73)$ & 0\\
  \hline
  \multirow{2}{*}{$\simeq 45$} & \multirow{2}{*}{0.15} & \multirow{2}{*}{10} & 100 & $42 \: (7.6)$ & $0.15 \: (0.13)$ & $10 \: (9.0)$ & 1 \\
  &&& 500 & $42 \: (7.7)$ & $0.15 \:(0.13)$ & $10 \: (9.0)$ & 2\\
  \hline
  \multirow{2}{*}{$\simeq 45$} & \multirow{2}{*}{0.6} & \multirow{2}{*}{10} & 100 & $44 \:(15)$ & $0.54  \:(0.21)$ & $9.9  \:(0.91)$ & 0\\
  &&& 500 & $44  \:(15)$ & $0.56 \:(0.21)$ & $ 9.9  \:(0.89)$ & 0\\
  \hline
  \multirow{2}{*}{$\simeq 45$} & \multirow{2}{*}{0.3} & \multirow{2}{*}{5} & 100 & $42  \:(12)$ & $0.30  \:(0.19)$& 5.0  \:(21) & 5\\
  &&& 500 & $42  \:(13)$ & $0.31  \:(0.20)$& $5.0  \:(23)$ & 6\\
  \hline
  \multirow{2}{*}{$\simeq 45$} & \multirow{2}{*}{0.3} & \multirow{2}{*}{20} & 100 & $44  \:(10)$ & $0.26  \:(0.14)$& $20  \:(1.4)$ & 0\\
  &&& 500 & $45  \:(10)$ & $0.29  \:(0.15)$& $20  \:(1.4)$ & 0\\
 \hline
\end{tabular}}
\caption{Inference results by \textit{flimo} on several populations simulated under the Ricker model, for different parameter values and different computational effort. The exact values of $r^*$ are $exp(3.8), exp(3.8)/2 \text{ and } 2exp(3.8)$. This value of $exp(3.8)$ is used in \cite{wood_statistical_2010}.
}
\label{table:ricker}
\end{minipage}
\end{center}
\end{table*}

\section{Discussion}\label{discussion}

%avantages de flimo
Concerning the point inference of parameters, \textit{flimo} has the advantage of being considerably faster than the other methods with comparable accuracy: up to 5,000 times faster for the g-and-k example, and up to 100 times faster on the Wright-Fisher model. Obviously, the efficiency of an algorithm strongly depends on its implementation, especially on the programming languages used. The algorithm \textit{flimo} is implemented in Julia, a language known for its good numerical performances. The other algorithms used here are implemented either in C++ for \textit{MCMC} and \textit{wABC}, or in Python with \texttt{Numpy} for \textit{compareHMM}, which are comparably efficient programming languages \citep{aruoba_comparison_2015}. This ensures that its computation speed is an intrinsic property of the \textit{flimo} algorithm.
Like in ABC methods, the choice of the summary statistics is important. However, \textit{flimo} seems to be less sensitive than the ABC implementations tested here. This lower sensitivity allows to select some summary statistics that can be calculated quickly, thus decreasing the optimization time of \textit{flimo}. This is illustrated by the use of the Wasserstein distance which is necessary to have good estimates of the parameters of g-and-k distributions using the ABC method, while it only slows down \textit{flimo} with low benefit on the accuracy of the results.
The behaviour of \textit{flimo} in high dimension (in terms of number of parameters) is the purpose of a further investigation (Supplementary File 1), based on a toy example from \cite{li_extending_2017}. It seems realistic to infer a few dozens of parameters.

%choix de l'algorithme d'optimisation

%The same is true for the use of the Beta with spikes distribution used by the \textit{compareHMM} algorithm, and which similarly only slows down \textit{flimo} because of the complexity of its implementation.

%autres approches
\textit{Flimo} also allows to obtain suitable empirical estimates of the distribution of the inferred parameters, but at the cost of its computation efficiency. Obtaining a distribution of parameters rather than a point value is relevant, in particular for estimating the uncertainties of these estimated parameters. An interesting strategy could be to implement some hybrid methods, using \textit{flimo} to circumscribe a region of interest in the search space, which could then be used as a fine prior for a Bayesian method. 

%limites de flimo
As with ABC, the model-based approach of \textit{flimo} has its limitations. In particular, the models selected by the user bias the analysis, since this prior choice influences the conclusions that will be drawn from the algorithm's results \citep{csillery_approximate_2010}. However, \textit{flimo} allows for the comparison of several different scenarios, provided that the same summary statistics are used for the different models.
The way in which \textit{flimo} works does, however, imply some constraints that other simulation-based methods do not have. First of all, the models studied must verify some properties. On the one hand, it is necessary to be able to calculate efficiently the quantiles of the concerned distributions. This is not always the case, as for Beta distributions. In these cases, it is advantageous to approximate the distribution, for example by a normal distribution. Moreover, \textit{flimo} needs that the states of the model do not large discontinuities, otherwise the optimization algorithms tend to fail. For example, \textit{flimo} is not suitable for building a most likely phylogenetic tree.
In addition, the usual constraints of optimisation problems apply here: there must not be many local minima for the considered objective function, that should ideally be convex. Unlike Bayesian methods which can explore the parameters space widely, \textit{flimo} tends to revolve around a single area of the search space, which depends on the initial conditions of the optimization. It is therefore necessary in practice to perform several inferences to ensure good convergence. Resampling methods, like jackknife, can be used to make the inference more robust.
In general, it does not seem possible to prove the convergence of the point estimators established by flimo to the theoretical parameters of the model. This is mainly due to the fact that summary statistics do not have a priori any particular properties, and therefore matching summary statistics from data and simulations does not prove that the parameter values are similar. A simple example of a calculation where convergence can be shown is given in Supplementary File 2. This illustrates the fact that the \textit{flimo} approach (both in the simulation and minimisation process) does not in itself induce an asymptotic bias.

%conclusion
 %Point estimators are obtained in an extremely efficient way in terms of computation time and with good accuracy.
 This study shows that \textit{flimo} is a solid alternative to other inference methods, especially ABC, with a simple implementation which, thanks to the Julia and R packages that we developed. It can easily be adapted to a wide range of models form different fields, for example in population dynamics or population genetics.
 By rethinking the role of randomness in stochastic model inference problems, \textit{flimo} makes it possible to use efficient deterministic gradient-based optimization algorithms to infer the parameters of probabilistic models whose likelihood or moments are intractable. All these qualities make \textit{flimo} a particularly simple and effective method of inference.

\backmatter

\section{Statement and Declarations}

\subsection{Competing interests}

 This work was supported by the Alpalga project (ANR-20-CE02-0020). The authors have no relevant financial or non-financial interests to disclose.

\subsection{Data and Code availability}

All the data and scripts are available on \url{https://metabarcoding.org/flimo}.

\section{Author contributions statement}

All authors conceived the algorithm. S.M., C.G. and E.C. wrote the manuscript. E.O. and D.P. contributed to writing the manuscript. S.M. developed the packages and performed computational experiments with assistance and guidance from C.G., E.C. and E.O. C.G. and E.C. supervised the project.

\section{Acknowledgments}

We thank Pierre Pudlo and Adeline Leclercq-Samson for their help on some theoretical aspects of the paper, and Simon Boitard for sharing with us his work on the Wright-Fisher model.

Some of the computations presented in this paper were performed using the GRICAD infrastructure (\url{https://gricad.univ-grenoble-alpes.fr}), which is supported by Grenoble research communities.


%\begin{appendices}

%\section{Section title of first appendix}\label{secA1}




%\end{appendices}


\bibliography{reference}% common bib file

\section*{Supplementary Information}

%\section*{Supplementary material}

\setcounter{table}{0}
\renewcommand{\thetable}{S\arabic{table}}%
\setcounter{figure}{0}
\renewcommand{\thefigure}{S\arabic{figure}}%

%\subsection*{Estimate of the parameters of g-and-k distributions}

\begin{figure}[!ht]
 \centering
 \includegraphics[width=0.95\linewidth]{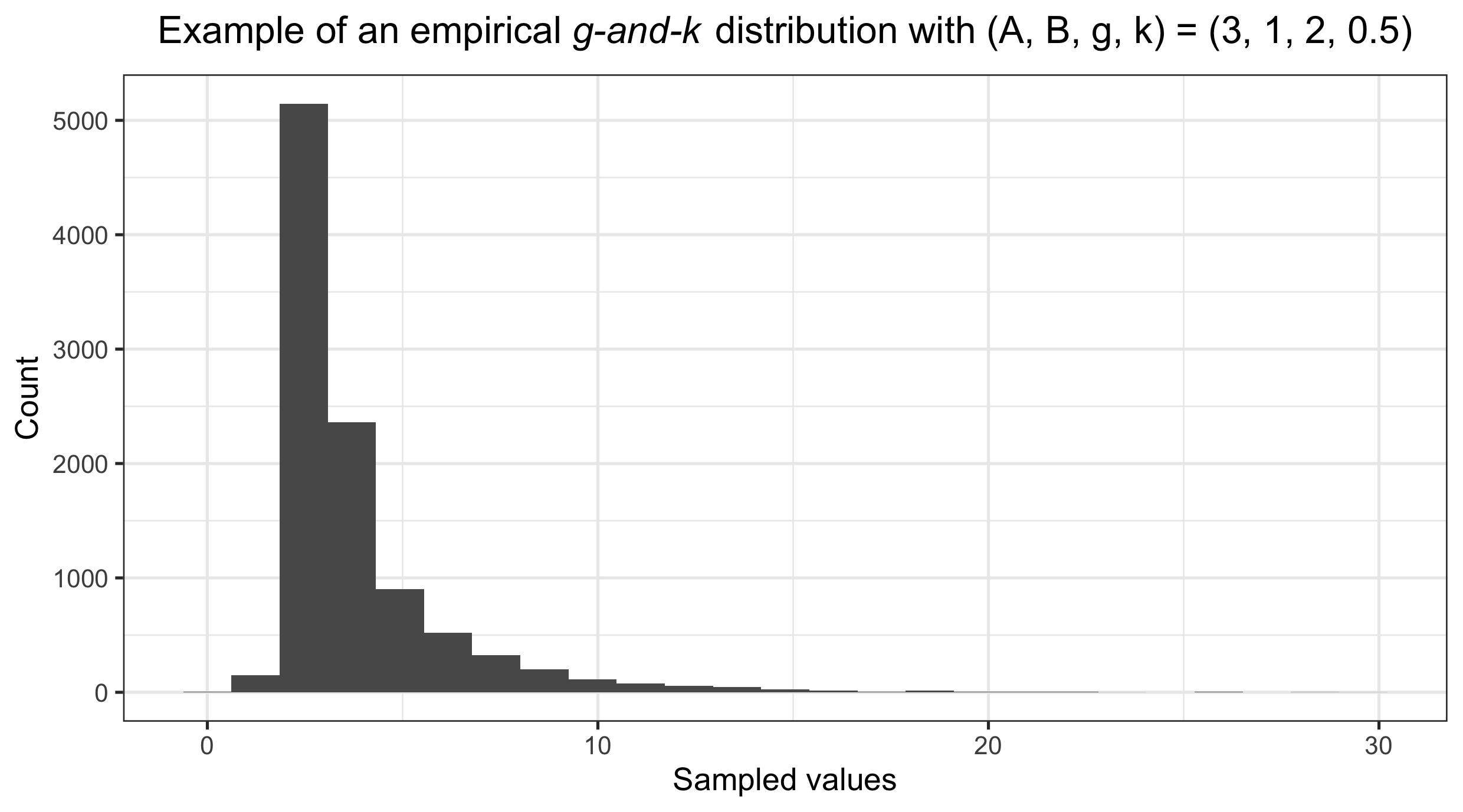}
 \caption{Empirical histogram of a g-and-k distribution for parameters $(A, B, g, k) = (3, 2, 1, 0.5)$.}
 \label{fig:gk_hist}
\end{figure}

\begin{figure}[!ht]
 \centering
 \includegraphics[width=0.95\linewidth]{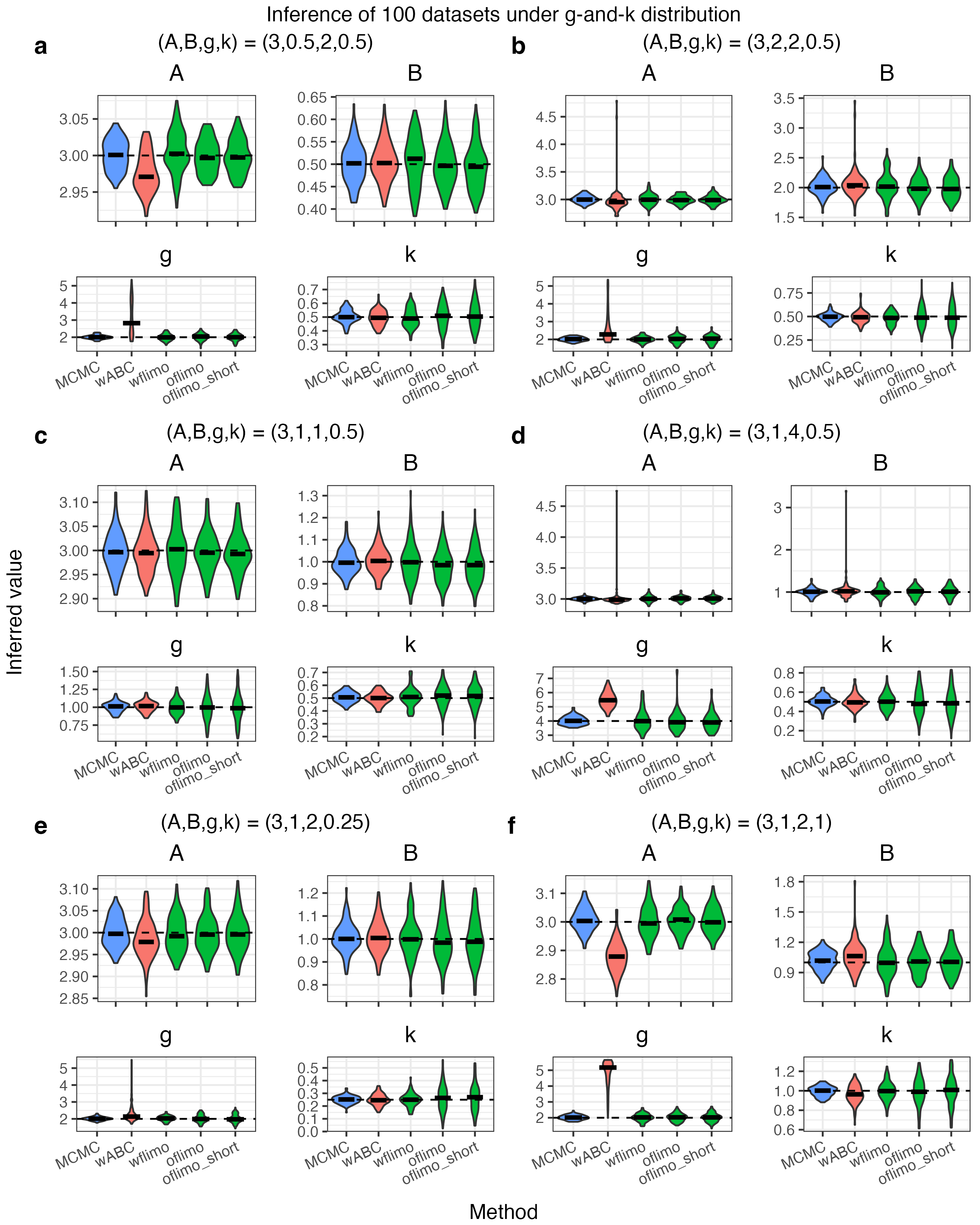}
 \caption{Estimate of the parameters of g-and-k distributions for different parameter sets. The \textit{oABC} and \textit{wABC-short} methods have not been included due to their poor performance on the main example. In the case where $g=4$ (panel d), i.e. when the peak of the g-and-k distribution is sharper, some \textit{flimo} inferences tend to overestimate $g$. An attempt to improve this inference by jackknifing was made. For the cases studied, the median of one hundred inferred values with jackknifing is closer to the theoretical value. The average of the inferred parameters is a weaker estimator, due to a certain number of outliers. The inference results of $g$ by \textit{wABC} are biased on panels d and f. This is not the case for \textit{flimo}. We did not try to correct the outliers isolated for $A$ and $B$ for the \textit{wABC} method (panels b and d in particular). The times were not reported because the different inference methods were not used under the same conditions.}
 \label{fig:gk_suppl}
\end{figure}

\clearpage

\begin{figure}[!ht]
 \centering
 \includegraphics[width=0.95\linewidth]{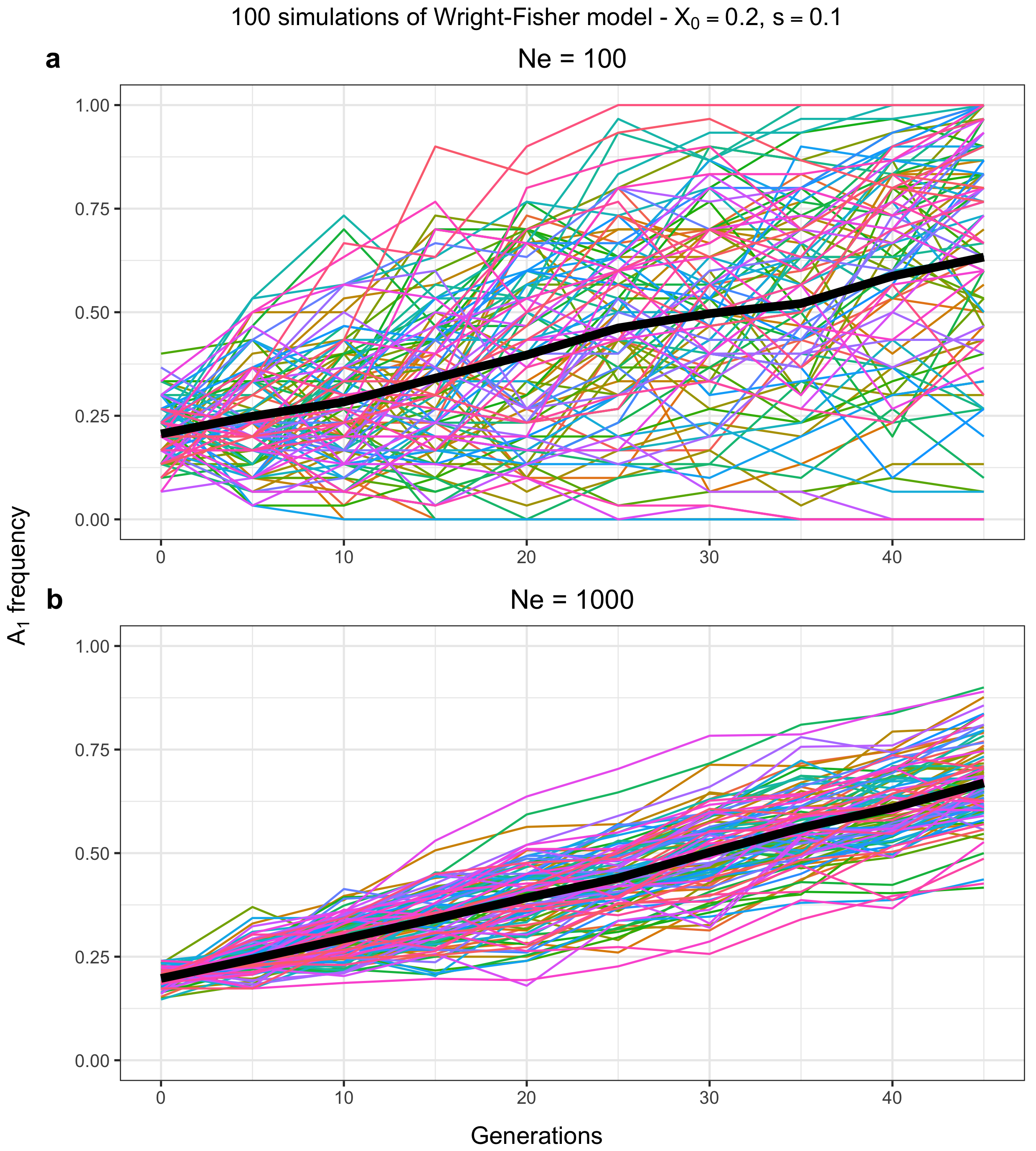}
 \caption{One hundred simulated data sets for $s = 0.1$ and two population sizes (\textbf{Panel a}: $N_e = 10^2$; \textbf{Panel b}: $N_e = 10^3$). Thick black line is the average $A_1$ proportion at each sampled time. \textbf{Panel a}: the dispersion from one simulation to another is important with a non-negligible probability of allele fixation.}
 \label{fig:wf_simu}
\end{figure}

%\subsection*{Estimation of a Wright-Fisher selection value - alternative scenarios}

\begin{table}[ht]
\begin{center}
\begin{minipage}{0.99\textwidth}
\resizebox{\textwidth}{!}{%
\begin{tabular}{ccccccccc}
\toprule%
 Selective & &\textit{compareHMM}& \multicolumn{6}{@{}c@{}}{\textit{flimo}} \\ \cmidrule{4-9}
value & Criterion & \textit{Bws} & \multicolumn{2}{@{}c@{}}{\textit{Binomial}} & \multicolumn{2}{@{}c@{}}{\textit{Bws}}& \multicolumn{2}{@{}c@{}}{\textit{NG}}\\ \cmidrule{4-5} \cmidrule{6-7} \cmidrule{8-9}
&&&10& 200 & 10& 200 & 10 & 200 \\
 \hline
 \multirow{5}{*}{$s = 0.01$} & Correlation &0.998&0.86 & 0.89&0.85&0.89&0.88&0.90\\
 \cmidrule{2-5} \cmidrule{6-7} \cmidrule{8-9}
 & Median difference &0.0& $3.2 \:10^{-3}$ & $2.5 \:10^{-3}$&$-5.0 \:10^{-4}$ & $2.8 \:10^{-3}$ & $4.3 \:10^{-4}$ & $9.5 \:10^{-4}$ \\
 &$\widehat{s}_\text{flimo} - \widehat{s}_\text{compareHMM-Bin}$&(0 \%) & (32\%) & (25\%)&(-5.0\%) & (28\%) & (4.3\%) & (9.5\%) \\ 
 \cmidrule{2-5} \cmidrule{6-7} \cmidrule{8-9}
 &Seconds by &\multirow{2}{*}{1.4}&\multirow{2}{*}{0.038} &\multirow{2}{*}{0.76}&\multirow{2}{*}{0.34}&\multirow{2}{*}{6.4}&\multirow{2}{*}{0.013}&\multirow{2}{*}{0.10}\\
&inference &&&&&&&\\
 \hline
 \multirow{5}{*}{$s = 1$} &Correlation &0.98&0.75 & 0.72&0.65&0.69&0.64&0.69\\
 \cmidrule{2-5} \cmidrule{6-7} \cmidrule{8-9}
&Median difference &$-6.7\:10^{-3}$& $-2.1\:10^{-2}$ &$-1.5\:10^{-2}$& $-3.0\:10^{-2}$&$-1.8\:10^{-2}$& $-2.5\:10^{-2}$ & $-2.2\:10^{-2}$ \\ 
 &$\widehat{s}_\text{flimo} - \widehat{s}_\text{compareHMM-Bin}$&(-0.67\%) & (-2.1\%) & (-1.5\%)& (-3.0\%) & (-1.8\%) & (-2.5\%) & (-2.2\%) \\ 
 \cmidrule{2-5} \cmidrule{6-7} \cmidrule{8-9}
&Seconds by &\multirow{2}{*}{3.0}&\multirow{2}{*}{0.015} &\multirow{2}{*}{0.29}&\multirow{2}{*}{0.22}&\multirow{2}{*}{3.8}&\multirow{2}{*}{0.015}&\multirow{2}{*}{0.12} \\
&inference &&&&&&&\\
 \hline
\end{tabular}}
 \caption{Estimation of a Wright-Fisher selection value - alternative scenarios. Inference results, based on 100 simulated data sets with $N_e = 10^3$ and $s \in \{0.01, 1\}$, using \textit{compareHMM} or different implementations of the \textit{flimo} method. Three quantities are presented: the Pearson correlation coefficient, the median of the difference between the values inferred by the \textit{flimo} or \textit{compareHMM-Bws}, and \textit{compareHMM-Bin} methods, and the median of the computation times. For several inferences with $s = 0.01$, there is however a systematic overestimation of \textit{flimo} compared to \textit{compareHMM-Bin}. Note that in the latter case, the mean of the estimated $s$ by \textit{flimo}is closer to the simulated value $s= 0.01$ than with \textit{compareHMM-Bin}. The large deviation from the theoretical value is due to important random fluctuations under these simulation conditions. For $s=1$, the mean values inferred by \textit{compareHMM} are more distant from the true value than those inferred by \textit{flimo}.}
\label{table:wfscenariobis}
\end{minipage}
\end{center}
\end{table}

\clearpage

\subsubsection*{Supplementary File 1: Performance of \textit{flimo} for a high dimension problem}

We adapt here the toy example from \cite{li_extending_2017}. This basic example allows to evaluate the limits of the inference algorithms. In \cite{li_extending_2017}, the aim is to illustrate the performance in the construction of the posterior distribution for different ABC implementations. In our case, it is more a question of testing the limits in numbers of parameters reasonably usable by \textit{flimo}. The model considered is given by the equation \ref{hd_model}.

\begin{align}
    y &\sim \mathcal{N}_p(\theta, \Sigma) \quad p \geq 2 \label{hd_model} \\
    \text{with } y &= (y_1, \dots, y_p)^T \nonumber \\
    \theta &= (\theta_1, \dots, \theta_p)^T \nonumber \\
    \Sigma &= diag(\sigma_0, \dots, \sigma_0) \nonumber
\end{align}

We use the prior chosen by the authors to determine the initial condition of the inference (equation \ref{hd_prior}). As in their study, we fix $\sigma_0 = 1$ and $b= 0.1$.

\begin{align}
    &\theta \sim \mathcal{N}(0, A) \text{ with } A = diag(100, 1, \dots, 1) \label{hd_prior} \\
    &\text{and then transforming } \theta_2 \leftarrow \theta_2 + b\theta_1^2 - 100b \nonumber
\end{align}

The data consist of a single observation $y_{obs} = (10, 0, \dots, 0)^T$ and the summary statistics are the data vector itself: $s(y) = y$. The distance between the summary statistics is the Euclidean distance. The objective function for \textit{flimo} is therefore given by the equation \ref{hd_j}. To evaluate \textit{flimo}, we choose $n_{sim} = 10$.

\begin{align}
    &J(\theta) = \sqrt{\sum_{i=1}^p(\overline{s(y_\theta)}_i - s(y_{obs})_i)^2} \label{hd_j}
\end{align}

where $\overline{s(y_\theta)}$ is the mean of the $n_{sim}$ simulations produced by \textit{flimo}.\\

In our case, \textit{flimo} does not construct a posterior distribution. Therefore the evaluation criterion chosen is not a Kullback-Leibler divergence for the distribution of $(\theta_1, \theta_2)$ but the Euclidean distance between the data and the inferred values, for the first two components (equation \ref{hd_crit}).

\begin{align}
   Error(\hat{\theta}) = (y_{obs, 1}-\hat{\theta}_1)^2 + (y_{obs,2}-\hat{\theta}_2)^2 \label{hd_crit}
\end{align}

The values of $p$ tested range from 2 to 50. Table \ref{table:hd} shows the results. The median inference time seems to be proportional to $p^a$ with $2 \leq a \leq 3$ and longer extreme cases are common. The error criterion is stable as $p$ increases.

\begin{table}[ht]
\begin{center}
\begin{minipage}{\textwidth}
\resizebox{\textwidth}{!}{%
\begin{tabular}{ccccc}
\toprule%
 p & Mean Error & Maximum Error & Median Time & Maximum Time\\
 \hline
 2 & 0.44 & 0.98 & 0.0049 & 0.13 \\
 5 & 0.42 & 0.84 & 0.019 & 0.29 \\
 10 & 0.40 & 1.0 & 0.086 & 4.1 \\
 20 & 0.40 & 1.4 & 0.40 & 16 \\
 30 & 0.40 & 1.3 & 1.2 & 47 \\
 40 & 0.34 & 0.97 & 2.4 & 85 \\
 50 & 0.38 & 0.91 & 9.8 & 162\\
 \hline
\end{tabular}}
 \caption{Inference accuracy and computation time (in seconds) as a function of the number of parameters $p$ of the multivariate normal model, for $n_{sim} = 10$, for 100 replicates.}
\label{table:hd}
\end{minipage}
\end{center}
\end{table}

\clearpage

\subsubsection*{Supplementary File 2: Assessment of convergence in a toy example}

In some very simple cases, it is possible to show analytically that the parameters inferred by \textit{flimo} are convergent estimators of the model parameters. In this example, the observed data are drawn in a normal distribution (equation \ref{ex_data}). The summary statistics are the empiral mean and variance (equation \ref{ex_ss}).

The goal is to infer $\theta^* = (\mu^*, \sigma^*)$. In practice, we want to show that the estimators built by \textit{flimo} converge to the empirical moments $(\overline{y}, \sigma_y^2)$. In this case, it is of course matching with the moment estimators.

\begin{align}
    Y^{obs} &= (y_1, \dots, y_n) \sim \mathcal{N}(\mu^*,\sigma^*) \: iid, \quad n \geq 2 \label{ex_data} \\
    \overline{y} &= Mean(Y^{obs}) \quad ; \quad \sigma_y^2 =  Var(Y^{obs}) \label{ex_ss}
\end{align}

In the following, $\theta = (\mu, \sigma),\: \mu \in \mathbb{R},\: \sigma > 0$. Let $m \geq 2$ be the number of simulations to generate. There is no need for $n = m$. Let $z_\theta(q) = \mu + \sigma \sqrt{2} \text{erf}^{-1}(2q-1), \: q \in ]0,1[$ be the quantile function of the distribution $\mathcal{N}(\mu, \sigma)$. The quantiles used for the simulations are fixed: let $Q = (q_1, \dots, q_m) \sim \mathcal{U}([0,1]) \: iid$ and $simQ :\theta \mapsto (z_\theta(q_1), \dots, z_\theta(q_m)) $ be the simulation function. \textit{simQ} entries are independent draws in $\mathcal{N}(\mu, \sigma)$ by inverse transform sampling. The chosen objective function to minimize to find optimal parameters is given by equation \ref{ex_J}.

\begin{align}
    J_Q(\theta) &= \left(\overline{y} - \overline{simQ(\theta)}\right)^2+\left(\sigma_y^2 - Var(simQ(\theta)) \right)^2 \label{ex_J} \\
    \text{where } \overline{simQ(\theta)} &= \frac{1}{m} \sum_{i=1}^m z_\theta(q_i) = \frac{1}{m} \sum_{i=1}^m \mu + \sigma \sqrt{2} \text{erf}^{-1}(2q_i-1) \nonumber \\
    &= \mu + \frac{\sigma \sqrt{2}}{m} \sum_{i=1}^m \text{erf}^{-1}(2q_i-1) = \mu + \sigma  . \overline{e(Q)} \nonumber \\
    \text{with } \overline{e(Q)} &= \frac{\sqrt{2}}{m} \sum_{i=1}^m \text{erf}^{-1}(2q_i-1) \nonumber \\
    \text{and } Var(simQ(\theta)) &= \frac{1}{m} \sum_{i=1}^m \left(z_\theta(q_i) - \overline{simQ(\theta)} \right)^2 = \frac{1}{m} \sum_{i=1}^m z_\theta(q_i)^2 - (\mu + \sigma  . \overline{e(Q)})^2 \nonumber
\end{align}

The computation with the unbiased estimator of the variance (normalized by $\frac{1}{m-1}$) does not lead to a close form of the solution. The formula is developed in equation \ref{ex_Jclose}.

\begin{align}
    \frac{1}{m} \sum_{i=1}^m z_\theta(q_i)^2 &= \frac{1}{m} \sum_{i=1}^m (\mu + \sigma \sqrt{2} \text{erf}^{-1}(2q-1))^2 = \mu^2 + \sigma^2 \overline{e^2(Q)} + 2 \mu \sigma \overline{e(Q)} \nonumber \\
    \text{with } \overline{e^2(Q)} &= \frac{2}{m} \sum \text{erf}^{-1}(2q_i-1)^2 \nonumber \\
    \text{so } Var(simQ(\theta)) &=\mu^2 + \sigma^2 \overline{e^2(Q)} + 2  \mu \sigma \overline{e(Q)} - (\mu + \sigma  . \overline{e(Q)})^2 \nonumber \\
 &= \sigma^2 \left(\overline{e^2(Q)} - \overline{e(Q)}^2\right) \nonumber \\
    J(\theta) &= \overline{y}^2 + \mu^2 + \sigma^2 \overline{e(Q)}^2 + 2 \mu \sigma \overline{e(Q)} - 2\overline{y} (\mu + \sigma \overline{e(Q)}) + \nonumber \\
    &\quad \sigma_y^4 + \sigma^4 \left(\overline{e^2(Q)} - \overline{e(Q)}^2\right)^2 - 2 \sigma_y^2 \sigma^2 \left(\overline{e^2(Q)} - \overline{e(Q)}^2\right) \label{ex_Jclose}
\end{align}

Let $\widehat{\theta} = \mathrm{argmin} J(\theta)$. Its explicit value can be obtain by equation \ref{ex_theta}.

\begin{align}
    \frac{\partial J}{\partial \mu} (\theta) &= 2\mu - 2 \overline{y} + 2 \sigma \overline{e(Q)} \nonumber \\
    \frac{\partial J}{\partial \sigma} (\theta) &= 2 \sigma \overline{e(Q)}^2 + 2 \mu \overline{e(Q)} - 2 \overline{y} \overline{e(Q)} + 4 \sigma^3 \left(\overline{e^2(Q)} - \overline{e(Q)}^2\right)^2 \nonumber \\
    & \quad - 4 \sigma^2_y \sigma \left(\overline{e^2(Q)} - \overline{e(Q)}^2\right) \nonumber \\
    \text{Solving } \frac{\partial J}{\partial \mu} (\widehat{\theta}) &= 0 \text{ and } \frac{\partial J}{\partial \sigma} (\widehat{\theta}) = 0 \text{ we get } \widehat{\mu} = \overline{y} - \widehat{\sigma} \overline{e(Q)} \text{ and then } \nonumber \\
    \widehat{\sigma} &= \frac{\sigma_y}{\sqrt{\left(\overline{e^2(Q)} - \overline{e(Q)}^2\right)}} \text{ and } \quad \widehat{\mu} = \overline{y} - \sigma_y \frac{\overline{e(Q)}}{\sqrt{\left(\overline{e^2(Q)} - \overline{e(Q)}^2\right)}} \label{ex_theta}
\end{align}

We can compute the limit when $m \rightarrow + \infty$. Recall that

$$\overline{e(Q)} = \frac{\sqrt{2}}{m} \sum_{i=1}^m \text{erf}^{-1}(2q_i-1) \quad \text{and} \quad \overline{e^2(Q)} = \frac{2}{m} \sum \text{erf}^{-1}(2q_i-1)^2$$

with $(\sqrt{2} \text{erf}^{-1}(2q_i-1))_i$ being iid draws in a distribution $\mathcal{N}(0,1)$. So $\overline{e(Q)}$ is a convergent estimator of the mean of $\mathcal{N}(0,1)$: $\lim_{m\to\infty} \overline{e(Q)} = 0$. In the same way, $\overline{e^2(Q)} - \overline{e(Q)}^2$ is a convergent estimator of the variance of $\mathcal{N}(0,1)$, so $\lim_{m\to\infty} \overline{e^2(Q)} - \overline{e(Q)}^2 = 1$. To conclude,

$$\lim_{m\to\infty} \widehat{\mu} = \overline{y} \quad , \quad \lim_{m\to\infty} \widehat{\sigma} = \sigma_y$$

\end{document}